\documentclass[%
 reprint,
 amsmath,amssymb
,aps
prb,
]{revtex4-1}

\usepackage{graphicx}% Include figure files
\usepackage{dcolumn}% Align table columns on decimal point
\usepackage{bm}% bold math
\usepackage{tabularx, booktabs}
\usepackage{braket}
\usepackage{wrapfig}
\usepackage[colorlinks=true]{hyperref}
\usepackage{bbold}

\renewcommand{\vr}{\textbf{r}}

\newcommand{\cG}{{\cal G}}
\newcommand{\cU}{{\cal U}}
\newcommand{\Tr}{{\mathrm{Tr}}}
\newcommand{\iw}{{i\omega}}
\newcommand{\iv}{{i\nu}}

\newcommand{\w}{{\omega}}
\newcommand{\p}{{\partial}}
\newcommand{\ra}{{\rightarrow}}
\newcommand{\ve}{{\varepsilon}}
\newcommand{\vR}{\mathbf{R}}

\begin{document}

\preprint{APS/123-QED}

\title{Diatomic molecule as a testbed for combining DMFT with electronic structure methods such as $GW$ and DFT}

\author{Juho Lee}
%\affiliation{Physics Department, XYZ University.}%Lines break automatically or can be forced with \\
\author{Kristjan Haule}%
\affiliation{%
Department of Physics \& Astronomy, Rutgers University, Piscataway, NJ 08854-8019, USA
}%

\date{\today}% It is always \today, today,

\begin{abstract}
  We implemented combination of DMFT and $GW$ in its fully
  self-consistent way, one shot $GW$ approximation, and quasiparticle
  self-consistent scheme, and studied how well these combined methods
  perform on H$_2$ molecule as compared to more established methods
  such as LDA+DMFT. We found that most flavors of $GW$+DMFT break down
  in strongly correlated regime due to causality violation. Among
  $GW$+DMFT methods, only
  the self-consistent quasiparticle $GW$+DMFT with static
  double-counting, and a new method with causal double-counting, correctly recover the atomic limit at large H-atom  separation. While some flavors of $GW$+DMFT improve the
  single-electron spectra as compared to LDA+DMFT, the total energy is best
  predicted by LDA+DMFT, for which the exact double-counting is known,
  and is static.
\end{abstract}

\maketitle

\section{Introduction}

In last couple of decades, many theories have been developed to tackle
the problem of strong correlations in systems where conventional
methods based on the density functional theory
(DFT)~\cite{kohn-hohenberg1964,kohn-sham1965} encounter difficulties.
Kinetic energy and electronic interaction are comparable in those
materials, and electrons display a mixed behavior between particles
and waves. This makes the problem complicated even for low-energy lattice
models with limited number of degrees of freedom. Such
delocalization-localization interplay of electron is the key to the
numerous interesting phenomena in condensed matter physics including
tunable magnetism, colossal magnetoresistance, heavy fermion
behavior, high-Tc superconductivity and metal insulator transition.

Among many theoretical developments, the dynamical mean-field theory
(DMFT)~\cite{georges-kotliar1992,kotliar1996rev} has been very
successful, and due to its non-perturbative nature it was able to
describe the phenomena of the first order metal insulator transition
even in its most simplistic form of the single band Hubbard
model.~\cite{kotliar1996rev} Due to the flexibility of the DMFT, it
has been extended in many directions, adding bosonic bath to treat
long range spin~\cite{si1996prl,kajuter1996rutgers} or coulomb interactions~\cite{chitra2000prl}, extending the
range of correlations from a site to clusters~\cite{maier2005rmp} to address the
issue of momentum space differentiation, and finally it has been
combined with DFT to become more realistic~\cite{kotliar2006rev}. This
combination of DFT and DMFT (DFT+DMFT) has been very successful in
describing materials with open $d$ and $f$ shells both for their
spectral properties, as well as computing total energy~\cite{amadon2006prl} and
free energy~\cite{birol2015prl} of crystal phases. Recently, DMFT has also been
successfully applied to molecules~\cite{lin2011prl,weber2014pnas}.

Hedin's $GW$ approximation \cite{hedin1965,onida2002rmp} is a
many-body perturbative technique, which approximates the self-energy
by the lowest order diagram in the screened Coulomb interaction:
$\Sigma=-GW$.  As opposed to the ground state nature of DFT, where
the gaps of the Kohn-Sham spectrum have no physical meaning, in $GW$
approximation the target is the single particle Green's function and
therefore the single particle excitation spectrum of $GW$ is expected to be
a better prediction than  the Kohn-Sham spectrum.
%
%While eigenvalues given by DFT do not have a direct physical
%interpretation except for the highest occupied value and the band gaps
%are in poor agreement with experiment, the $GW$ spectral is understood
%as particle/hole excitation peaks and predicts better bandstructures.
%
Since $GW$ is a diagrammatic method and DMFT can also be expressed in
the diagrammatic form, the combination of the GW and DMFT ($GW$+DMFT)
was proposed~\cite{sun2002prb,biermann2003prl} as a possibly better
alternative to DFT+DMFT.
%
%One of the advantages of $GW$+DMFT over LDA+DMFT is that it is free
%from double-counting issue because the intersection of correlation
%energy of $GW$ and DMFT is obvious diagrammatically.  
%
Furthermore, the momentum dependence of $GW$ self-energy is expected
to complement the local nature of the DMFT, in particular when
the DMFT locality is enforced in less localized basis, such as in the
basis of Wannier orbitals.~\cite{tomczak2012prl,miyake2013prb,brouet2013prl}.  

Most of $GW$ calculations for solid-state systems in practice rely
on one-shot $GW$ scheme (generally denoted as $G_0W_0$). In this
scheme, the $GW$ self-energy is computed only once and the
non-interacting Green's function $G_0$ is obtained from the Kohn-Sham
 DFT spectrum~\cite{louie1986prb}.
 The one-shot $GW$ method has been successful for many materials with
 weak to moderate electronic correlation, giving a very good
 approximation for bandgaps in semiconductors~\cite{aryasetiawan1998gw}.  To remove the
 dependence on the DFT spectrum through $G_0$, the scheme called
 quasiparticle self-consistent QS-$GW$ was developed
 \cite{faleev2004prl,schilfgaarde2006prl,kotani2007prb} where the
 non-interacting $G_0$ is determined in a self-consistent way from the
 $GW$ spectrum.

 While the success of the DFT+DMFT is now supported by the numerous
 applications to solid state systems, which are too many to review
 here, the GW+DMFT method is still in its infancy. Nevertheless,
 several calculations implementing some flavor of $GW$+DMFT have been
 reported recently, both for real
 materials~\cite{tomczak2012epl,sakuma2013prb,taranto2013prb,tomczak2014prb,choi2015}
 and for models~\cite{ayral2012prl,hansmann2013prl,ayral2013prb}.
 However a comprehensive test of numerous GW+DMFT schemes, and their
 appropriateness for calculating spectra or energy, is still
 lacking.

% Unfortunately, fully dynamical and self-consistent $GW$+DMFT has not
% been achieved yet except for model studies because of its demanding
% calculation cost. This prevents us from investigating the effect of
% the level of self-consistency which is the remaining issue of
% $GW$+DMFT that needs to be answered.
%
% Furthermore, a new approach has been recently developed, combining QS$GW$ and DMFT
% \cite{choi2015,taranto2013prb,tomczak2014prb}.
% This scheme significantly reduces the calculation cost by replacing 
% the dynamical $GW$ part with an effective static Hermitian operator, 
% which is computed in such a way that 
% the eigenvalues of the effective Hamiltonian
% matches the quasiparticle energies of the fully dynamical one.
% Since QS$GW$+DMFT retains the same basic structure as
% LDA+DMFT, it can be naturally implemented with widely used conventional DMFT calculation package.
% Although the basic idea has been established, there is a subtle issue of double-counting which needs to be tested in detail.

 Small molecules have served as very good test beds to investigate
 electronic structure methods. For example, Lin \emph{et al.}
 \cite{lin2011prl} applied DMFT to hydrogen chain (H$_n$) where
 they found the cluster DMFT produces comparable accuracy to density
 matrix renormalization group (DMRG). Our previous study of
 LDA+DMFT on H$_2$ molecule \cite{juholee2015prb} demonstrated that
 single-site DMFT with a good choice of local orbitals and exact
 double-counting method gives extremely precise total energy, and also
considerably improves the spectra, as compared to LDA.

Here we perform a comprehensive test of various flavors of GW+DMFT,
from fully self-consistent $GW$+DMFT to $G_0 W_0$+DMFT and
QS$GW$+DMFT. We compute the total energy and the spectra of H$_2$
molecule for all these methods, and compare them to DFT+DMFT and the
exact solution.

We find that the strongly correlated regime is very challenging for
most of GW+DMFT methods, and most of them fail due to causality
violation, which has not been properly addressed before.  The only
exception proposed before is the quasiparticle self-consistent method, i.e.,
QS$GW$+DMFT. The latter recovers the correct atomic limit only when
combined with the static double-counting, in which case it gives
comparable spectra and energy to the results of LDA+DMFT.
We propose here a new causal double-counting scheme,
which works better than other GW+DMFT schemes in the strongly correlated regime.

Close to the equilibrium volume, which can be characterized as the
weakly to moderately correlated regime, most of GW+DMFT schemes considerably
improve the spectra, as compared to LDA+DMFT. However, the total
energy of GW+DMFT rivals LDA+DMFT total energy only in the fully self-consistent
version, which however breaks down in the correlated regime. The
quasiparticle method with static double-counting, which performs
well in the correlated regime, does not give very accurate total
energy in the moderately correlated regime, as it is not derivable
from a functional.

Our study demonstrates that the QS$GW$+DMFT, in combination with 
static double-counting, is a promising direction for computing
spectra of correlated systems both in moderate to strongly correlated
regime. On the other hand, the GW+DMFT methods tested in this work, do
not rival LDA+DMFT in predicting the total energy of the system.

% In this paper, we report calculation results for 
%  various flavors of $GW$+DMFT from SC-$GW$+DMFT to
% $G_0 W_0$+DMFT to QS$GW$+DMFT on the H$_2$ molecule.
% Since the simple nature of H$_2$ molecule allows us to perform fully self-consistent $GW$+DMFT calculation within first-principle, we see the effect of self-consistency in detail. 
% Moreover, the causality, which is not guaranteed in the $GW$+DMFT scheme, is closely investigate.
% For QS$GW$+DMFT, two different double-counting schemes are implemented and compared to each other.

\section{Methods}

\subsection{Approximations derived from a Functional: DMFT, LDA+DMFT, and GW+DMFT}

Let us start by refreshing the Baym-Kadanoff (BK)
formalism~\cite{baym_kadanoff,baym1962}, 
which provides the functional 
of the Green's
function $G$ 
\begin{equation}\label{bk}
\Gamma[G] = \Tr\log G-\Tr((G^{-1}_{0}-G^{-1})G)+\Phi_v[G], 
\end{equation}
that is stationary for the exact Green's function $G$, and gives the
grand potential, when evaluated on the exact Green's function $G$ (for
details see also \cite{kotliar2006rev}).
Here $G_0$ is the non-interacting Green's function
$G_0^{-1} = [\iw+\mu-\nabla ^2 +V_{ext}(\vr)]\delta(\vr,\vr')$ and
$V_{ext}$ is the potential due to nuclei.
The last term $\Phi_v[G]$ is the so-called Luttinger-Ward (LW)
functional, which is the sum of all skeleton diagrams constructed by 
the Green's function $G$ and the Coulomb repulsion 
$v(\vr,\vr')=\frac{1}{|\vr-\vr'|}$. 
The derivative of the LW functional with respect to $G$ gives the
exact self-energy of the system 
\begin{equation}\label{lw_exact}
\frac{\delta \Phi_v[G]}{\delta G}=\Sigma.
\end{equation}
The stationarity of the functional $\Gamma[G]$ 
at the exact $G$ ($\delta \Gamma/{\delta G} =0$) is ensured by 
the Dyson equation
\begin{equation}\label{dyson_eq}
G^{-1} -[\iw+\mu-\nabla ^2 +V_{ext}(\vr)]\delta(\vr,\vr') +\Sigma=0.
\end{equation}
The functional $\Phi_v[G]$ is diagrammatically known, but its
evaluation is extremely difficult due to fermionic minus sign problem~\cite{prokofiev2006pre}.
Nevertheless, this formalism is extremely useful because many good
approximations can be devised by approximating $\Phi$ rather than
$\Gamma$, and such approximations were shown to be
conserving~\cite{baym_kadanoff,baym1962}.

One can classify these approximations into two classes: those that
truncate correlations in the real space, and those that truncate in
the space of Feynman diagrams. In LDA the functional $\Phi$ is
truncated in real space so that exchange and correlations are local to
a point in 3D space, i.e., each point in 3D space is mapped to an
independent auxiliary problem of electron gas. In the DMFT the
functional $\Phi$ is also truncated in real space, but the locality is
constraint to a site on the lattice, which is mapped to an auxiliary
problem of quantum impurity.

On the other hand, in Hartree-Fock and $GW$ theories, the truncation
is done in the space of Feynman diagrams but the complete space
dependence of the self-energy is kept. The $GW$+DMFT can then be seen
as the hybrid between these two classes of approaches, as it truncates
Feynman diagrams only for the long range part of the correlations,
while the short-range correlations can be exactly accounted for by the DMFT.

% In principle, the exact form of the functional $\Phi_v[G]$ is unknown.
% Therefore, we try to solve problems with 
% methods such as DMFT, LDA+DMFT and $GW$+DMFT.
% These frameworks approximates LW functional $\Phi_v[G]$
% by certain few terms or sub-classes of an infinite sum of diagrams which can
% capture the most important features of the system.
% The $\Phi$-derivable methods are often called conserving approximation
% because the solution always preserve conserving quantities 
% restricted by symmetries of the given system.

%which means that the solution $G$ of \eqref{dyson_eq} automatically satisfies 
%the equations of motion of physical observables 
%such particle number, momentum and energy.
%One of the advantages of it is that the expectation values of observables are independent of
%the way it is calculated. For example, energies evaluated from Galitskii–Migdal
%formula and Baym-Kadanoff free-energy should be the same.
%Furthermore, since the solution of $\Phi$-derivable methods is where 
%the free-energy is extrimized, the energy around the solution 
%is not sensitive with respect to small error of Green's function.

\subsubsection{DMFT}

In the DMFT method, the locality of correlations is explored and the
LW-functional is truncated so that it is a functional of the local
Green's function ($G_{loc}$) only, i.e., it contains all skeleton Feynman diagrams
that are constructed from $G_{loc}$ and interaction $v$, and all
diagrams that are outside this range, are neglected. In real material calculation, 
the DMFT method is defined only, once the projector to the
local Green's function is specified. In this work, we will use the
real space projectors, defined by
\begin{align}\label{projection}
G^{\mathbf{R}}_{loc} &=\hat P G \equiv \sum_{LL'} \ket{\chi^{L}_{\mathbf{R}}} \braket{\chi^{L}_{\mathbf{R}}|G|\chi^{L'}_{\mathbf{R}}}\bra{\chi^{L'}_{\mathbf{R}}}.
\end{align}
where $\{\ket{\chi^{L}_{\mathbf{R}}}\}$ is a local orbital set
centered on a given nucleus at $\mathbf{R}$, and $L$ is an orbital
index. In the single-orbital DMFT case, which we will test in this
work, no sum over $L$ is needed.

The DMFT LW-functional is then 
\begin{equation}\label{lw_dmft}
\Phi^{\mathrm{DMFT}}[G]=\sum_{\mathbf{R}}\Phi_v[G_{loc}^{\mathbf{R}}].
\end{equation}
and the functional $\Phi_v[G_{loc}]$ has the same dependence on
$G_{loc}$ as the exact functional $\Phi_v[G]$ on $G$, except that it
has finite range. This is because any Feynman diagram of arbitrary
topology that is contained in exact $\Phi_v[G]$ is also contained in
approximate $\Phi_v[G_{loc}]$. In solid state applications of DMFT,
the interaction
$v$ has to be replaced by the screened interaction $\cU$ due to the
fact that many degrees of freedom are being removed from
consideration. Screening in molecules is negligible, hence we can
safely take $\cU=v$ for the molecular systems.

To compute $\Phi_v[G_{loc}^{\mathbf{R}}]$ the system is mapped to a
quantum impurity model, for which $\cG_{imp}=G_{loc}^{\mathbf{R}}$ so
that the exact solution of the impurity problem $\Phi_v[\cG_{imp}]$
delivers the desired LW functional. The DMFT self-energy is then obtained
from the auxiliary impurity self-energy, as required by the
Baym-Kadanoff formalism:
\begin{align}\label{embedding}
 \Sigma^{\mathrm{DMFT}}&= \frac{\delta \Phi^{\mathrm{DMFT}}}{\delta G}=
 \sum_{\mathbf{R}}\frac{\delta G^{\mathbf{R}}_{loc}}{\delta G}\frac{\delta \Phi_v[G^\vR_{loc}]}{\delta G^{\mathbf{R}}_{loc}}\nonumber\\
 &=\sum_{\mathbf{R},LL'} \ket{\chi^L_{\mathbf{R}}}\Sigma^{\mathbf{R},LL'}_{imp}\bra{\chi^{L'}_{\mathbf{R}}}\equiv \hat E \Sigma_{imp}
 \end{align}
%  &=\sum_{i} \ket{\chi_{i}}\bra{\chi_{i}}\Sigma^{i}_{imp}\equiv \hat E \Sigma_{imp}
%where $G$ is assumed to be a matrix element.
where we define the embedding $\hat E$. This embedding process is the inverse operation of the projector $\hat P$, mapping the self-energy of the auxiliary impurity back into the Hilbert space of the original system. %We mention in passing that from now on, we drop the site index $\mathbf{R}$ in the local quantities $\cG^{\mathbf{R}}_{imp}$ and $\Sigma^{\mathbf{R}}_{imp}$ because the two sites of H$_2$ are symmetric.

\subsubsection{LDA+DMFT}

%Even though LDA+DMFT is not a main topic of this paper, we also
%present the LDA+DMFT formalism described in our previous work
%\cite{juholee2015prb} because the comparison between $GW$+DMFT and
%LDA+DMFT is one of our interests. The application of LDA+DMFT to H$_2$
%molecule was very successful for the total energy accuracy and the
%dissociation process while the spectral result was not as accurate as
%the total energy. Therefore, the LDA+DMFT result must be a good
%reference data as well as the exact solution from the configuration
%interaction method.

In the LDA+DMFT formalism, the LW functional is
approximated by the combination of LDA and DMFT functional
\begin{align}\label{lw_ldmft}
\Phi^{\mathrm{LDA+DMFT}}&[G]=E_v^{H}[\rho]+E_v^{X}[\rho]+E_v^{\mathrm{LDA},C}[\rho]\nonumber\\
&+\sum_{\mathbf{R}}(\Phi_{v}[G^{\mathbf{R}}_{loc}]-\Phi^{DC}[G^{\mathbf{R}}_{loc}])
\end{align}
where the double-counting (DC) correction $\Phi^{DC}$ subtracts
the intersection of the two approximations.
The Hartree, exchange, and correlation functional of the LDA are
\begin{align}
&E_v^{H}[\rho]=\frac{1}{2}\int_{\vr\vr'} \rho(\vr)v(\vr-\vr')\rho(\vr')\label{lw_H}\\
&E_v^{X}[\rho]=-\frac{1}{2}\sum_\sigma\int_{\vr\vr'}\rho^\sigma(\vr,\vr')v(\vr-\vr')\rho^\sigma(\vr',\vr) \label{lw_X}\\
&E_v^{\mathrm{LDA},C}[\rho]=\int_{\vr}\rho(\vr)\varepsilon_v^{\mathrm{LDA},C}[\rho(\vr)]. \label{lw_C}
\end{align}
We used the Vosko-Wilk-Nusair (VWN) parametrization for the LDA
correlation energy~\cite{corr_vw}. 
%Note that we use here the exact
%exchange Eq.~\eqref{lw_X} rather than semilocal exchange, as
%typically used in condensed matter applications.

% is the exchange functional (in
% paramagnetic cases such as H$_2$, $\rho^{\sigma}=\rho/2$) instead of
% the Slater-exchange approximation which is commonly used one for DFT
% calculation. We found that the effect of non-local exchange in small
% molecules is quite significant that the local Slater-exchange scheme
% does not produce a precise result.

When applying the DMFT approximation on the LDA functional, (or, applying
the LDA approximation on the DMFT functional) we get the following
expression for the "exact" double-counting:
\begin{align}\label{dc_ldmft}
\Phi^{DC}[G^{\mathbf{R}}_{loc}]=E_{v}^{H}[\rho^{\mathbf{R}}_{loc}]+E_{v}^{X}[\rho^{\mathbf{R}}_{loc}] +
E^{\mathrm{LDA},C}_{v}[\rho^{\mathbf{R}}_{loc}]
\end{align}
where $\rho^{\mathbf{R}}_{loc}$ is the projected local density
$\rho^{\mathbf{R}}_{loc}= G^\vR_{loc}(\tau=0^-)$ at the site $\mathbf{R}$.
% In the case of H$_2$, since the screening effect must be negligible,
% we can approximate $\cU$ as $\cU\sim v$. Therefore, the each term of 
% the double-counting can be easily obtain by plugging $\rho_{imp}(\vr,\vr')$
% instead of $\rho(\vr,\vr')$ in \eqref{lw_H}-\eqref{lw_C}.
%the impurity interaction is simply $\cU_{imp}=\bra{\chi_i\chi_i}U_C(\vr-\vr')\ket{\chi_i\chi_i}$ since 
As shown in Ref.~\cite{juholee2015prb}, this LDA+DMFT with the exact
double-counting leads to very accurate results in H$_2$ molecule. Note
that such exact double-counting can also be extended to solid state
systems, but in this case one needs to replace Coulomb interaction $v$
with the screened interaction $\cU$ in the functional
Eq.~\ref{dc_ldmft}. ~\cite{haule2015prl}

\subsubsection{$GW$+DMFT}

The $GW$ self-energy is the first order approximation of 
screened interaction $\Sigma^{GW}=-GW$ (we follow here the
sign convention of the imaginary time formalism).
The screened interaction $W=v/(1-Pv)$ is
approximated by the RPA polarization $P=2GG$
where the factor 2 is for spin degrees of freedom.
From \eqref{lw_exact}, its functional form can be written as
\begin{align} \label{lw_gw}
\Phi_v^{GW}[G]&=-\sum_{n=1}^{\infty}
                \frac{1}{2n}\Tr[(2G v G)^{n}]=\frac{1}{2}
                \mathrm{Tr~log}(1-2G v G)
\end{align}
where the first term (n=1) corresponds to 
the exchange functional $E_v^{X}[\rho]$.

Just like in LDA+DMFT, LW functional of $GW$+DMFT is a combination of $GW$ functional
augmented with the DMFT functional for the local degrees of freedom, i.e.,
\begin{align}\label{lw_gdmft}
\Phi^{GW+\mathrm{DMFT}}[G]=&E_v^{H}[\rho]+\Phi_v^{GW}[G]\nonumber\\
&+\sum_{\mathbf{R}}(\Phi_v [G^{\mathbf{R}}_{loc}]-\Phi^{DC}[G^{\mathbf{R}}_{loc}]).
\end{align}
The double-counting of $GW$+DMFT is obtained by applying the DMFT
approximation on the $GW$ functional, leading to
\begin{eqnarray}
\Phi^{DC}[G^\vR_{loc}]=
E_v^H[\rho_{loc}^\vR]+\Phi_v^{GW}[G_{loc}^\vR]
\nonumber\\
=E_v^H[\rho_{loc}^\vR]+ \frac{1}{2}\Tr \log(1-2 G^\vR_{loc} v G_{loc}^\vR).
\label{lw_localgw}
\end{eqnarray}

%The Dyson equation $\delta \Gamma[G]/\delta G=0$ then gives 
%\begin{equation} \label{dyson_gdmft}
%G^{-1}=[\iw+\mu -H^H(\vr)]\delta(\vr-\vr')-\Sigma(\vr\vr';\iw)
%\end{equation}
%%^{GW+\mathrm{DMFT}}
%where we include the Hartree potential $v^{H}(\vr)=\int d\vr' v(\vr-\vr') \rho(\vr')$ in the Hartree operator, not in the self-energy part:
%\begin{gather}
%H^H(\vr)=[-\nabla ^2 +v_{ext}(\vr)+v^{H}(\vr)]\\
%\Sigma(\vr\vr';\iw)= \Sigma^{GW}(\vr\vr';\iw)+\hat E(\Sigma_{imp}(\iw)-\Sigma_{DC}).
%\end{gather}

To converge the $GW$+DMFT equations, we implemented the following
steps:
\begin{itemize}
\item[(1)] Starting with an initial non-interacting Green's function
  $G=G_0$, we construct 
\begin{gather}
P(\tau)=2G(\tau) G(-\tau)\\
W(\iv)=v/(1-P(\iv)v)
\end{gather}
where all variables above are general matrices.\\
\item[(2)] The $GW$ self-energy is given by:
\begin{equation}\label{gw_self}
\Sigma^{GW}(\vr\vr';\tau)=-G(\vr\vr';\tau)W(\vr\vr';\tau)
\end{equation}
\item[(3)] From \eqref{lw_localgw}, we see that the double-counted $GW$
  contribution to the self-energy is :
\begin{equation}\label{gwloc_self}
\Sigma^{GW}_{DC}(\tau)=\frac{\delta \Phi_v^{GW}[G^\vR_{loc}]}{\delta
  G^\vR_{loc}} =- G^\vR_{loc}(\tau)W_{loc}(\tau)
\end{equation}
where the local components are computed by
\begin{gather}
 G^\vR_{loc}=\hat P^\vR G \\ 
 P_{loc}(\tau)=2 G^\vR_{loc}(\tau) G^\vR_{loc}(-\tau)\\
 W_{loc}(\iv)=v/(1-P_{loc}(\iv) v).
 \end{gather}
\item[(4)] Electron density is given by $\rho= G(\tau=0^{-})$.
\\
\item[(5)] Next we calculate the Hartree potential
\begin{equation}
V^{H}(\vr)=\frac{\delta \Phi^{H}[\rho]}{\delta \rho(\vr)}=\int d\vr' v(\vr-\vr') \rho(\vr').
\end{equation}
\item[(6)] (\emph{DMFT loop}) Using local Green's function at each
  site $G^\vR_{loc}=\hat P_\vR G$ and the interaction $v$, 
we construct the auxiliary impurity model, which delivers the
impurity self-energy
\begin{equation}
\Sigma_{imp}= \frac{\delta \Phi_v[G^\vR_{loc}]}{\delta G_{loc}^\vR}
\end{equation}
% where $\Delta(\iw)$ is hybridization function, which serves as the
% dynamical mean-field of the auxiliary impurity. 
% The detail of DMFT loop will be discussed later in \ref{discuss_hyb}.\\
\item[(7)] Putting together GW, DMFT and DC, the total self-energy is obtained by
\begin{align}
\Sigma=& \Sigma^{GW}+\sum_{\mathbf{R},LL'} \ket{\chi^L_{\mathbf{R}}}(\Sigma^{\vR,LL'}_{imp}-\Sigma_{DC}^{\vR,LL'})\bra{\chi^{L'}_{\mathbf{R}}}
\end{align}
The double-counting is $\Sigma_{DC}(\iw)=
V_{loc}^H+\Sigma_{DC}^{GW}(\iw)$, with $\Sigma_{DC}^{GW}$ evaluated in
(3),  and the local Hartree $V^{H}_{loc}$ is 
\begin{align}
 V^{H}_{loc}&= \frac{\delta \Phi^{H}[\rho_{loc}]}{\delta \rho_{loc}}\nonumber\\
&=\sum_{LL'} \ket{\chi^{L}_{\mathbf{R}}} \braket{\chi^{L}_{\mathbf{R}}| V^{H}[\rho_{loc}] |\chi^{L'}_{\mathbf{R}}}\bra{\chi^{L'}_{\mathbf{R}}}.
\end{align}
\item[(8)] From Dyson equation \eqref{dyson_eq}, the total Green's
  function is given by
\begin{equation} \label{dyson_gdmft}
G^{-1}=[\iw+\mu+\nabla^2-V_{ext}-V^{H}]\delta(\vr-\vr')-\Sigma(\vr\vr';\iw),
\end{equation}
where the chemical potential is determined by enforcing charge
neutrality, i.e.,
\begin{equation}
\int d\vr \rho(\vr)= \int d\vr G(\vr,\vr;\tau=0^{-}) =Z_{nuclei}
\end{equation}
\item[(9)] For {\emph{Fully self-consistent GW+DMFT}}, go back to (1).
  All variables are updated until self-consistency is reached.
  For {\emph{G$_0$W$_0$+DMFT}}, go to (4). Therefore the
  $GW$ self-energy \eqref{gw_self} and its local counterpart
  \eqref{gwloc_self} do not change over the iterative process while
  the impurity self-energy, total density and the Green's function are
  updated.
\end{itemize}

\subsection{Quasiparticle self-consistent $GW$+DMFT and its double-counting}

First, let us discuss the quasiparticle self-consistent $GW$ ~\cite{faleev2004prl,schilfgaarde2006prl,kotani2007prb}.
It is similar to $G_0W_0$ in that the polarization $P=2G^{\mathrm{QP}}G^{\mathrm{QP}}$ and the
self-energy $\Sigma=-G^{\mathrm{QP}} W^{\mathrm{QP}}$ are
computed from a free-particle Green's function
$G^{\mathrm{QP}}=1/(\w+\mu-H^{\mathrm{QP}})$, in which $H^{\mathrm{QP}}$ is
a Hermitian non-interacting Hamiltonian, which is however determined
in a self-consistent way from the GW spectra.

Refs.~\cite{faleev2004prl,schilfgaarde2006prl,kotani2007prb}  proposed
to solve the following quasiparticle equation
\begin{equation}\label{gw_pole}
\big[ -\nabla^2+V_{ext}+V^H+\mathrm{Re}\Sigma^{GW}(E_n)-E_n\big]\ket{\psi_n}=0.
\end{equation}
to determined the Hermitian quasiparticle
Hamiltonian with the form $\big[ H^{\mathrm{QP}}-E_n \big] \ket{\psi_n}=0$.

% Therefore, our goal is to construct an effective static Hamiltonian $H^{\mathrm{QP}}$ such that 
% it also generates the same eigenvalues $\big[ H^{\mathrm{QP}}-E_n \big] \ket{\psi_n}=0$.

Since the GW self-energy has a weak frequency dependence, we may
use
$\mathrm{Re}\Sigma^{GW}(\w)\approx 
\mathrm{Re}\Sigma^{GW}(0)+\frac{\p\mathrm{Re}\Sigma^{GW}(0)}{\p \w} \w=
\mathrm{Re}\Sigma^{GW}(0)+(1-Z^{-1})\w$
where the quasiparticle renormalization amplitude matrix is
\begin{equation}\label{qs_renorm}
Z^{-1}= \mathbb{1}-\frac{\p \mathrm{Re}\Sigma^{GW}(0)}{\p \w},
\end{equation}
which gives the following form for the quasiparticle Hamiltonian
\begin{equation}\label{Hqsgw}
H^{\mathrm{QP}}=Z^{1/2} \big[-\nabla^2+V_{ext}+V^H+\mathrm{Re}\Sigma^{GW}(0)\big] Z^{1/2}.
\end{equation}

Since the QS$GW$ procedure provides a static effective Hamiltonian
in which the $GW$ spectral information is encoded,
one might think that QS$GW$ Hamiltonian can simply replace 
the KS Hamiltonian of DFT+DMFT and yield better accuracy.
However, to implement QS$GW$+DMFT, there is a subtle issue concerning the double-counting between the $H^{\mathrm{QP}}$ and DMFT correlation. 

Since $H^{\mathrm{QP}}$ is constructed based on the real part of $GW$
self-energy, one may attempt to 
approximate $H^{\mathrm{QP}}+\Sigma(\iw)=H^{\mathrm{QP}}+\hat
E(\Sigma_{imp}(\iw)-\mathrm{Re}\Sigma^{GW}_{DC}(\iw))$.
But this self-energy obviously violates the causality condition as it
does not respect the Kramers-Kronig relation. 
In the following, we will introduce two
double-counting schemes, which obey the 
Kramers-Kronig relation.

\subsubsection{Static double-counting (SDC)}

In the simplest approach, we can take the double-counting as the
zero-frequency value of the local-GW self-energy, i.e.,
\begin{equation}\label{SDC}
\Sigma^{SDC} = V^{H}_{loc}+\mathrm{Re}\Sigma^{GW}_{DC}(\w=0)
\end{equation}
where $\Sigma^{GW}_{DC}$ is the exact DC given by Eq.  \eqref{gwloc_self}.
Recently, the combined method QS$GW$+DMFT, has been implemented for
real materials in Ref.~\cite{choi2015}, in which the static double-counting was
employed. 

We implement QS$GW$+DMFT as follows:
\begin{itemize}
\item[(1)] We start with initial values for $G^{\mathrm{QP}}$, $\rho(\vr)$ and $\Sigma_{imp}(\iw)$. 
We take their LDA counterparts $H^{KS}$,
$(G^{\mathrm{QP}})^{-1}=\iw+\mu-H^{KS}$ and $\rho^{\mathrm{LDA}}$.
For the impurity self-energy, we start with the local Hartree-Fock.

\item[(2)] Self-energies are constructed $\Sigma^{GW}=-G^{\mathrm{QP}}W^{\mathrm{QP}}$ and $\Sigma_{DC}^{GW}=-G^\mathrm{QP}_{loc}W^\mathrm{QP}_{loc}$
%@@@@
and then we obtain the double counting $\Sigma^{SDC}=V^{H}_{loc}+\Sigma_{DC}^{GW}(w=0).$

\item[(3)] Next, the Hartree potential is computed from the density
\begin{equation}
V^{H}(\vr)=\frac{\delta \Phi^{H}[\rho]}{\delta \rho(\vr)}=\int d\vr' v(\vr-\vr') \rho(\vr').
\end{equation}

\item[(4)] The quasiparticle Hamiltonian $H^{\mathrm{QP}}$ is computed
  using Eq.~\eqref{Hqsgw}, so that new $(G^{\mathrm{QP}})^{-1}=\iw+\mu-H^{\mathrm{QP}}$.

\item[(5)] The total Green's function is then given by
\begin{equation}\label{Gqsgwd}
G=\frac{1}{\iw+\mu-H^{\mathrm{QP}}-\hat E (\Sigma_{imp}-\Sigma^{SDC})}.
\end{equation}
\item[(6)] The density and the chemical potential are computed from
  $\rho(\vr)=G(\vr,\vr,\tau=0^{-1})$.

\item[(7)] (DMFT loop) from the local Green's function $G^\vR_{loc}=\hat
P^\vR G$ and the interaction $v$, the impurity solver calculates 
a new impurity self-energy: 
\begin{equation}
(G^\vR_{loc},v)~\ra~\Sigma_{imp}(\iw).
\end{equation}
\item[(8)] With updated variables $G^{\mathrm{QP}}$, $\rho$ and $\Sigma^{imp}$, go to the step (2).\\
\end{itemize}
% @@@@
The main difference between our QS$GW$+DMFT and that of Ref.~\cite{choi2015}
is that the quasiparticle $H^{\mathrm{QP}}$ in Ref.~\cite{choi2015} was calculated 
from QS$GW$ only, and the feedback of the DMFT self-energy on $H^{\mathrm{QP}}$ was
ignored. In our case, we recompute $H^{\mathrm{QP}}$ from the physical self-consistent
Green's function in every iteration.

\subsubsection{Dynamical double-counting (DDC)}

Although the static local GW term is subtracted, one can expect that
the spectral function is possibly over-renormalized because
$H^{\mathrm{QS}GW}$, which is renormalized by $GW$, is again
renormalized by the DMFT self-energy.

This is a very subtle issue for QS$GW$+DMFT because the dynamical
effects of $GW$ self-energy are incorporated in the static QS$GW$
Hamiltonian $H^{\mathrm{QP}}$, therefore we would like to subtract
the local part of this renormalization.

To overcome this problem, we first construct a \emph{non-local}
quasiparticle Hamiltonian $H^{\mathrm{QP}}_{nonloc}$ where not only
$GW$ but also the subtraction of the local $GW$ is incorporated. The
bandwidth of correlated orbitals of $H^{\mathrm{QP}}_{nonloc}$ must be
wider than $H^{\mathrm{QP}}$ because we \emph{unrenormalize} the local
$GW$ effect in $H^{\mathrm{QP}}_{nonloc}$. This widened band is then
corrected by the impurity self-energy, which is expected to be more
accurate than the local $GW$ self-energy.

In order to define the non-local quasiparticle Hamiltonian
$H_{nonloc}^{\mathrm{QP}}$, using the non-local $GW$ self-energy we write
\begin{equation}
\Sigma^{GW}_{nonloc}(\omega)=\Sigma^{GW}(\omega)-\Sigma^{GW}_{DC}(\omega),
\end{equation}
and we compute
\begin{equation}
\bar Z^{-1} = \mathbb{1} - \frac{\p \Sigma^{GW}_{nonloc}(0)}{\p \w}
\end{equation} 
and then we define
\begin{equation}
{H}_{nonloc}^{\mathrm{QP}}=\bar Z^{1/2}\big[ -\nabla+V_{ext}+V^{H}+\Sigma_{nonloc}^{GW}(0) \big]\bar Z^{1/2}.
\end{equation}
The above algorithm then has to be modified so that the step (5) in computing
the total Green's function uses
\begin{equation}
G=\frac{1}{\iw+\mu-H_{nonloc}^{\mathrm{QP}}-\hat E (\Sigma_{imp}-V^{H}_{loc})}.
\end{equation}
rather then \eqref{Gqsgwd}.

This DDC approach shares the basic idea of the scheme introduced by Tomczak in Ref.~\cite{tomczak1}, where he calculated the $H^{\mathrm{QP}}_{nonloc}$ along the real frequency. On the other hand, we implement the scheme based on the Matsubara formalism using linearization of the $GW$ self-energy.

\subsection{Causal double-counting scheme for $GW$+DMFT}
We propose here another type of double-counting for $GW$+DMFT, 
which we denote as \emph{causal double-counting} (CDC).
As will be clear in the III. A. 1, $GW$+DMFT with the exact double-counting (Eq.~\eqref{gwloc_self})
suffers a causality violation that does not allow $GW$+DMFT to work in the strongly correlated regime.
To avoid the causality breakdown, we introduce causal double-counting (CDC) functional
\begin{equation}\label{CDC}
\Sigma^{DC}=\hat P \Sigma^{GW}.
\end{equation}
and we will discuss why this double-counting scheme allows us to
avoid the causality issue in the III. A. 2.

One can notice that the CDC is not exact because this double-counting
contains the diagrams in which the degrees of freedom of the DMFT local orbitals
and the rest of the space interact through the screened interaction $W$, 
which is not contained in the DMFT self-energy.
Nevertheless, it allows $GW$+DMFT to work in the strongly correlated regime 
without violating the causality.

\subsection{Computational details}
In this work we use only the single site DMFT combined with LDA and
various flavors of GW. We use the same choice of the DMFT projector as
in our previous study of LDA+DMFT~\cite{juholee2015prb}, the linear
combination of the lowest two orbitals of H$_2^+$ cation,
$\ket{1s\sigma_g}$ (bonding) and $\ket{1p\sigma_u}$ (anti-bonding)
state. We denote them as the ``left" (L) and the ``right" (R)
localized orbital: \begin{align}\label{loc_orbital}
|\chi_L \rangle = \frac{1}{\sqrt{2}} ( |1s\sigma_{g}\rangle - |1p\sigma_{u}\rangle),\nonumber \\
|\chi_R \rangle = \frac{1}{\sqrt{2}} ( |1s\sigma_{g}\rangle + |1p\sigma_{u}\rangle).
\end{align}
This orbital set is a good choice for the DMFT projector because i)
they are well-localized at each atomic site, ii) they naturally
recover 1$s$ orbital (the ground state of $H$) on each site at large
atomic separation, iii) over 96\% of the electronic charge of the DMFT
solution is contained in these two states, which implies most of
correlation can be captured within the single site approach, and iv)
they do not explicitly depend on the self-consistent charge density.
The last condition is especially important for a stationarity of the
DMFT solution, given that we are extremizing the Luttinger-Ward
functional.

\begin{figure}[h!]
\centering{
\includegraphics[width=1.0\linewidth,clip=]{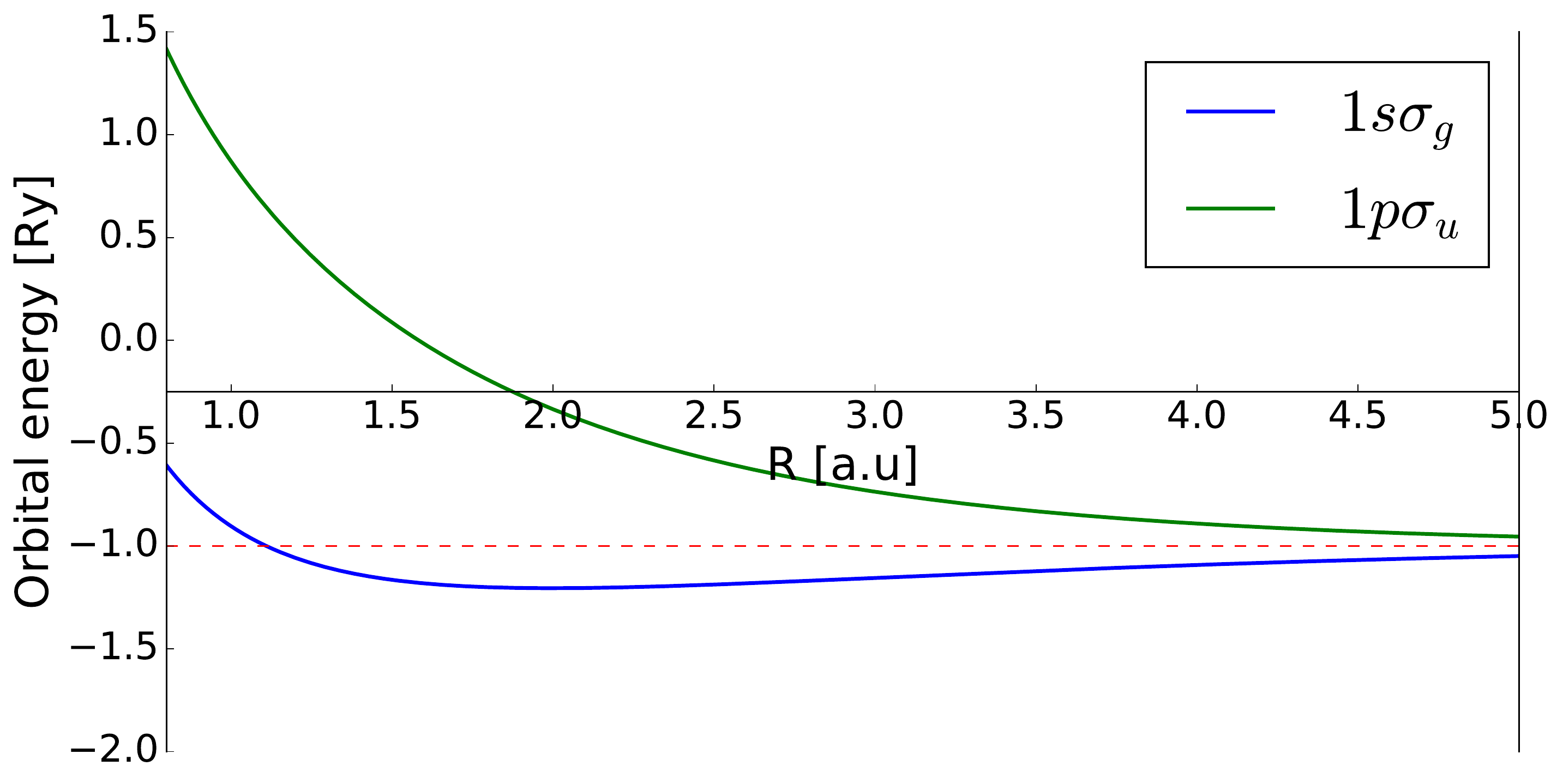}}
\caption{(Color online) The lowest two orbital energies of H$^{+}_2$ cation as a function of $R$, which are taken to define the DMFT projector (Eq.~\eqref{loc_orbital}).}
\label{fig0b}
\end{figure}

Since H$_{2}^{+}$ is a one-electron problem, the solution is achieved by solving 
the single-particle Schr{\"o}dinger's equation.
We follow a recursive approach (see Ref.~\cite{hadinger1989}) to solve H$^{+}_2$ cation and several lowest orbital energies are presented in Fig.~\ref{fig0b}.

The entire Hilbert space of H$_2$ is spanned by approximately 30
Gaussian orbitals (correlation-consistent basis set, cc-pVTZ
\cite{dunning1989jcp}). 
We want to emphasize that the $GW$ calculation in this study is
converged with respect to the size of the basis set, which is very challenging in solid state
applications, and this is another reason why such tests of GW+DMFT are
important and useful.

We evaluate the ground state energy of $GW$+DMFT schemes using the Galitskii-Migdal formula
\begin{equation}\label{GM_formula}
E=\Tr (H_0 \rho) + \frac{1}{2} \Tr (\Sigma G)
\end{equation}
where H$_0$ is the non-interacting part of Hamiltonian
$H_0=-\nabla^2+v_{ext}$, $\rho$ is the total electron density, and $G$ is
the total Green's function of the system.

The inverse temperature is set to be $\beta=1/k_B T=100Ry^{-1}$. Since
the orbital energy gap of H$_2$ is order of several $Ry$, this
temperature is sufficiently low and therefore describes the ground
state.

\section{Results and discussion} \label{results}

%\subsection{H$_2$ molecule as a strongly correlated system}

The H$_2$ molecule is a archetypical correlated system, often taken as
an example to demonstrate the failure of methods that use the single
slater determinant ansatz, such as Hartree-Fock and LDA. In the
dissociation limit, such methods predict delocalized ground state,
which never recovers correct atomic limit. $GW$ approximation, 
a many-body perturbative method, only slightly improves on LDA in this strongly
correlated limit as well.
The electronic correlations are only moderate around equilibrium
distance ($R=1.4$ a.u.), nevertheless none of these methods (LDA, HF
or GW) give an accurate total energy compared to the exact solution,
achieved by the configuration interaction (CI) method.

%half of the paper

In addition the prediction of the ionization energy (IE) from single
particle spectral function is a very good indicator of the quality of
the predicted single particle spectra within a given approximation. As
is known from the exact solution, the ionization energy is the
energy required to remove a single electron, i.e.,
\begin{align}\label{ie}
&\mathrm{H}_2+\mathrm{IE} ~\ra~ \mathrm{H}^{+}_2+e^{-}\nonumber \\
\Rightarrow ~&\mathrm{IE}=E({\mathrm{H}_2^{+}})-E({\mathrm{H}_2})
\end{align}
where $E({\mathrm{H}_2^{+}})$ is the ground state energy of H$_2^{+}$.
We computed the single particle Green's function in all tested
approaches, and checked how well they predict the position of the peak
in the spectral function corresponding to the IE energy.

\subsection{GW+DMFT}

\subsubsection{Total energy}
\begin{figure}[h!]
\centering{
\includegraphics[width=1.0\linewidth,clip=]{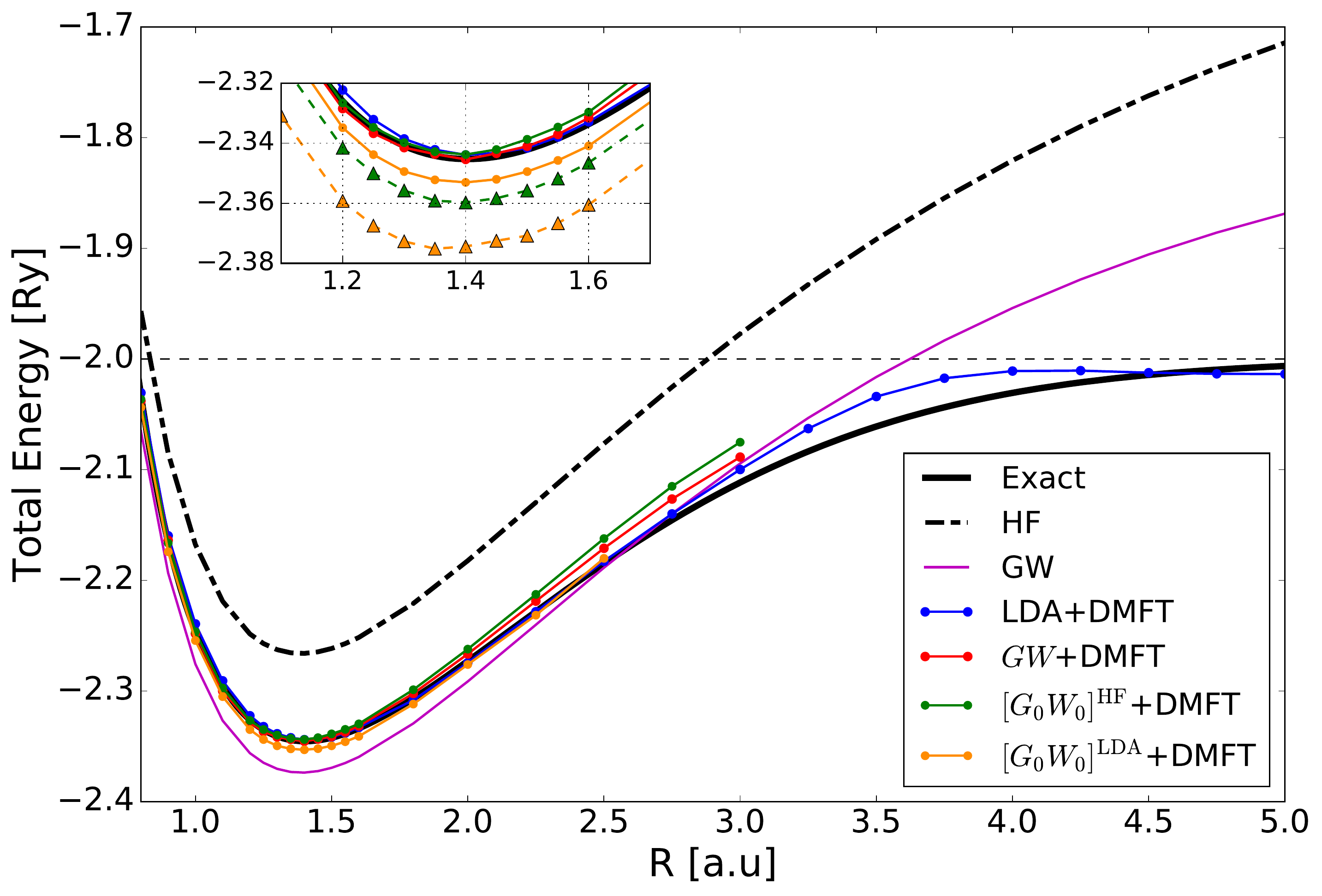}
}
 \caption{(Color online) The ground state energy of $GW$+DMFT versus $R$ presented with the HF, GW, LDA+DMFT and exact result for comparison. (\emph{inset}) We also added the total energy results of $G_0W_0$+DMFT without charge self-consistency (dashed lines).} \label{fig1}
\end{figure}

Fig.~\ref{fig1} show the total energy of several GW+DMFT methods and
compares it to the results of LDA+DMFT, HF, GW and the exact solution.
The LDA+DMFT results were already presented in our previous
work~\cite{juholee2015prb}, where we checked the accuracy of this
approximation, and we showed the importance of using the exact
double-counting within LDA+DMFT.
The accuracy of the predicted total energy within LDA+DMFT is
excellent, giving correct limit at large distance $R$ of
$-2.0$~Ry, and overall error below 1\%, with only exception around the
breking of the molecule ($R=3.5-4$), where non-local corrections
become important. At equilibrium distance, the error is less than
0.2\%.

As shown in Fig.~\ref{fig1} all methods tested here give better total
energy than Hartree Fock (3.5\% error) around the equilibrium distance
($R=1.4$). The self-consistent $GW$ gives error of approximately
1.3\%. Inclusion of the correlations captured by the DMFT improves the
total energy substantially, for example the fully self-consistent
$GW$+DMFT has an error of 0.3\% (very similar to 0.2\% in LDA+DMFT),
In $G_0W_0$+DMFT calculation, with $G_0$ being based on the
Hartree-Fock Hamiltonian ($[G_0W_0]^{\mathrm{HF}}$+DMFT), the accuracy
is almost as good as in $GW$+DMFT, while in
$[G_0W_0]^{\mathrm{LDA}}$+DMFT, the total energy is slightly less
precise.

% First of all, one can notice in Fig.~\ref{fig1} that the inclusion of
% the local DMFT self-energy enhances the total energy accuracy
% significantly. Around the equilibrium distance $R=1.4$, the HF and LDA
% approach gives a comparably huge error bar 3.5\% while the $GW$
% calculation, where the first order diagram of the screened interaction
% is taken into account, shows an acceptable accuracy (1.3\% errorbar).
% 
% The LDA+DMFT result in this regime is in an excellent agreement to the
% exact curve, giving the errorbar less than 0.2\%.
% 
% The fully self-consistent $GW$+DMFT (the red line in
% Fig.\ref{fig1}(a)), shows also an excellent agreement, 0.3\% of
% errorbar, which is just comparable to that of LDA+DMFT. For
% $G_0W_0$+DMFT calculation, when $G_0$ is based on Hartree-Fock
% Hamiltonian ($[G_0W_0]^{\mathrm{HF}}$+DMFT), its accuracy is almost
% the same as fully self-consistent one. For
% $[G_0W_0]^{\mathrm{LDA}}$+DMFT, the total energy becomes slightly less
% precise but is still good enough.

To show the effects of self-consistency of $GW$+DMFT in detail, we
display in the inset of Fig.~\ref{fig1} the $G_0W_0$+DMFT result at
different self-consistent level.  The dashed line  
shows $G_0W_0$+DMFT calculation where the
total electronic charge is not updated, which is the common practice
in solid state applications. In this approach, we first perform
self-consistent LDA (or HF) calculation, and then we fix the Hartree
potential $V^H$ and $GW$ self-energy
$\Sigma^{GW}=G_0^{\mathrm{LDA(HF)}}W_0^{\mathrm{LDA(HF)}}$ at the LDA (HF) level,
and we perform self-consistent DMFT calculation. Alternatively, when
the electronic charge is updated self-consistently on the $GW$+DMFT
charge, the accuracy of the total energy clearly improves.

It is interesting to note that the charge self-consistency has much
stronger effect on the total energy than the choice of the
non-interacting $G^0$ in $G^0W^0$. Both $G^0W^0$+DMFT methods are
quite close to the $GW$+DMFT results when the charge is updated, and
much worse when charge is fixed at the LDA/HF level. Perhaps this is
not very surprising, as the Hartree term contributes most to the total
energy. For total energy calculation, the charge self-consistency is then
much more important than the choice of $G_0$, despite the fact that
the $G_0$ based on LDA  is substantially worse (37\% error for IE)
than the HF spectra (2\% error for IE).

In Fig.~\ref{fig1} we could not continue $GW$+DMFT methods towards the
the atomic limit, most interesting correlated regime. The fully self-consistent
$GW$+DMFT and $G_0^{\mathrm{HF}}W_0^{\mathrm{HF}}$+DMFT break down around $R=3\,a.u$,
while $G_0^{\mathrm{LDA}}W_0^{\mathrm{LDA}}$+DMFT breaks down already at $R=2.5\,a.u.$.
The reason for such dramatic failure of $GW$+DMFT is the causality
violation, which we will address in the next section. This is one of
the most significant findings of our work, which shows that the
self-consistent $GW$+DMFT or $G_0W_0$+DMFT, when using exact
double-counting, have no future in addressing the problem of strong
correlations.

\subsubsection{Causality breakdown}
\label{discuss_hyb}
\label{result_hyb}

To solve the DMFT problem, and sum all local skeleton diagrams
$\Phi_v[G_{loc}^\vR]$ we construct an auxiliary impurity problem,
which has the same interaction $v$ as the original problem, and
$G_{loc}^\vR=\cG_{imp}$. Note that in solid-state systems we need to
renormalize interaction $v$ due to screening effects, which is not
needed here. Note also that this mapping of the local skeleton
diagrams to an impurity model is exact, and no further approximation
is made in this step. Furthermore, it is convenient to represent the
impurity Green's function in terms of proper and improper self-energy
( $\Sigma_{imp}$ and $\Delta$), i.e.,
\begin{eqnarray}
\cG_{imp}=\frac{1}{\w-\ve_{imp}-\Delta-\Sigma_{imp}}
\label{Impurity}
\end{eqnarray}
where  $\Sigma$ is the self-energy due to the Coulomb interaction,
while improper part $\Delta$ is due to the hybridization of this site with
the medium, and is therefore commonly referred to as the Weiss mean
field. The causality is violated if any of the three quantities
$\cG_{imp}$, $\Sigma_{imp}$, or $\Delta$ acquire positive imaginary
part at any frequency point on the real or imaginary axis.

We want to write the DMFT self-consistency condition
$\cG_{imp}=\hat{P}G$ in such a way that the Weiss mean field $\Delta$
from Eq.~\ref{Impurity} is expressed explicitly. To derive $\Delta$,
we will first eliminate the degrees of freedom which are
not corrected by the DMFT (the 30 Gaussian orbitals Hilbert space,
which has no overlap with the DMFT projectors). We will call this part
of the Hilbert space $r$. The part of the Hilbert space, which is
corrected by the DMFT will be denoted by $d$. In the second step, we
will extract the Green's function of a single site, which is needed by
the single site DMFT, and appears in the equation for hybridization $\Delta$.

We start with the Green's function of $GW$+DMFT from Eq.\ref{dyson_gdmft}:
\begin{eqnarray}
G =\left( {\omega+\mu-H^H-\Sigma^{GW}-\hat{E}(\Sigma_{imp}-\Sigma^{GW}_{DC})}\right)^{-1}
\end{eqnarray}
where we denoted $H^H=-\nabla^2+v_{ext}+v^H$. We next write it in the
block form, where $dd$ part of the matrix is
corrected by the DMFT, and the rest is not:
\begin{widetext}
\begin{eqnarray}
G=
\left(
\begin{array}{cc}
\left[\omega+\mu-H^H-\Sigma^{GW}\right]_{dd}-\Sigma_{imp}+\Sigma^{GW}_{DC} & -\left[H^H+\Sigma^{GW}\right]_{dr}\\
\\
-\left[H^H+\Sigma^{GW}\right]_{rd} & \left[\omega+\mu-H^H-\Sigma^{GW}\right]_{rr}
\end{array}
\right)^{-1}
\end{eqnarray}
We then eliminate the $r$ part of the matrix, so that the $G_{dd}$ becomes
\begin{eqnarray}
G_{dd} = \left[\mathbb{1} (\omega+\mu-\Sigma_{imp}+\Sigma^{GW}_{DC}) -
  (H^H+\Sigma^{GW})_{dd}-M_{dr} M_{rr}^{-1}M_{rd}\right]^{-1}
\end{eqnarray}
\end{widetext}
where we denoted
\begin{gather}
M_{dr(rd)}=[H^{H}+\Sigma^{GW}]_{dr(rd)}\nonumber\\
M_{rr}= [\omega+\mu-H^{H}-\Sigma^{GW}]_{rr}.
\end{gather}
We emphasized here that the $dd$ part of $G$ is still a matrix, in our
case $2\times 2$ for the two $H$ atoms. In solid state, the $dd$ part
would be an infinite matrix, containing the correlated
degrees of freedom, but written in real space.
In the second step we express the Green's function of a single site,
as needed by the DMFT.  We first define a matrix $S$:
%\begin{eqnarray}
%S \equiv (H^H+\Sigma^{GW})_{dd} + [H^{H}+\Sigma^{GW}]_{dr} [\omega+\mu-H^{H}-\Sigma^{GW}]_{rr}^{-1} [H^{H}+\Sigma^{GW}]_{dr}
%\end{eqnarray}
\begin{eqnarray}
S \equiv (H^H+\Sigma^{GW})_{dd} + M_{dr} M_{rr}^{-1} M_{rd}
\end{eqnarray}
so that 
\begin{eqnarray}
G_{dd} = \left[\mathbb{1} (\omega+\mu-\Sigma_{imp}+\Sigma^{GW}_{DC}) -S\right]^{-1}
\end{eqnarray}
and then the local Green's function becomes
\begin{widetext}
\begin{eqnarray}
G_{loc}=\frac{1}{\omega+\mu-\Sigma_{imp}+\Sigma_{DC}^{GW}-S_{11}-S_{12}\frac{1}{
  \omega+\mu-\Sigma_{imp}+\Sigma_{DC}^{GW}-S_{22}}S_{21}}
\label{Eq:47}
\end{eqnarray}
\end{widetext}

The crucial point is that in the correlated regime $\Sigma_{imp}$
becomes large (diverges) and therefore we can neglect the last term in
the denominator. Physically, this comes from the fact that in the
correlated regime the correlated sites (H-atoms) decouple, and the
effective hopping between them is thus cut-off by the appearance of
large local $\Sigma_{imp}$, and therefore DMFT is able to recover the
correct atomic limit. We thus have 
\begin{eqnarray}
  G_{loc}= \frac{1}{\omega+\mu-\Sigma_{imp} +\Sigma_{DC}^{GW}-S_{11}
  -O(\frac{1}{\Sigma_{imp}})}
\label{Eq:DMFT_loc}
\end{eqnarray}
and comparing Eq.~\ref{Eq:DMFT_loc} with Eq.~\ref{Impurity} revelas
\begin{eqnarray}
\Delta=&
\underbrace{ (M_{dr} \frac{1}{ (\w+\mu)\mathbb{1}_R- H^{H}_{rr}  -  \Sigma^{GW}_{rr}}M_{rd})_{11}}_{\equiv\Delta_{R}}
\nonumber\\
&+(\tilde \Sigma_{11}^{GW}- \tilde \Sigma_{DC}^{GW}) +O(\frac{1}{\Sigma_{imp}})\label{hyb:res}
\end{eqnarray}
where the tilde notation on the self-energy means 
$\tilde \Sigma(\w)=\Sigma(\w) - \Sigma(\infty)$,
and
$\varepsilon_{imp}=-\mu+H^{H}_{dd}+\Sigma^{GW}_{dd}(\infty)-{\Sigma^{GW}_{DC}}_{dd}(\infty)$.
Note that $\Sigma_{11}^{GW}=\hat P \Sigma^{GW}$.

Although we used in Eq.~\ref{Eq:47} the fact that $S$ is a $2\times 2$ matrix,
it is very easy to check that the resulting Eq.~\ref{hyb:res} is valid in
general, even in the solid state with infinite number of correlated
sites, a long as $\Sigma_{imp}$ is large, and sites decouple.

While the first term in Eq.~\ref{hyb:res} ($\Delta_R$) is always
causal, the second term is generally not, and its imaginary part can
have any sign. It has usually the non-causal sign, because the
$\Sigma^{GW}_{DC}$ tends to be larger than $\Sigma^{GW}_{dd}$.
As we will show in the section below, in the correlated
regime $\Delta_R$ becomes small, and then the hybridization becomes
non-causal.

Naively one would expect that $\Sigma^{GW}_{11}$ and
$\Sigma^{GW}_{DC}$ would cancel, but they do not, because the 
projected local self-energy $\Sigma^{GW}_{11}=\hat P \Sigma^{GW}$ is
\begin{equation}
\Sigma_{11}^{GW}=[-GW]_{11}=-G_{loc}W_{1111}-\sum_{r\ne 1}G_{rr}W_{1r1r}
\label{projected:sig}
\end{equation}
with the screened Coulomb interaction $W=v[1-Pv]^{-1}$. Here
$G_{11}=G_{loc}$. On the other hand, the
double-counted term $\Sigma_{DC}^{GW}$ is
\begin{equation}
\Sigma_{DC}^{GW}=-G_{loc}W_{loc}.
\end{equation}
with the screened local interaction defined by $W_{loc}=v[1-G_{loc} v
G_{loc}]^{-1}$. The two terms are then always different.

Note also that within LDA+DMFT, this problem does not occur, because
DC is static, and projected Kohn-Sham hamiltonian is also static,
hence causality can not be violated.

In the dissociation regime ($R>3.5\sim4.0$) the hopping between the
two H-atoms should vanish, and this can be achieved by diverging impurity
self-energy, so that the last term in the denominator of
Eq.~\ref{Eq:47}, vanishes. In this way we recover the exact atomic
limit. And indeed this is how LDA+DMFT achieves the exact atomic
limit. On the other hand $GW$+DMFT breaks down in this regime, and we
will show below that this is because $\textrm{Im}\Sigma_{11}^{GW}<\textrm{Im}\Sigma_{DC}^{GW}$.

\begin{figure}[h!]
\includegraphics[width=1.0\linewidth,clip=]{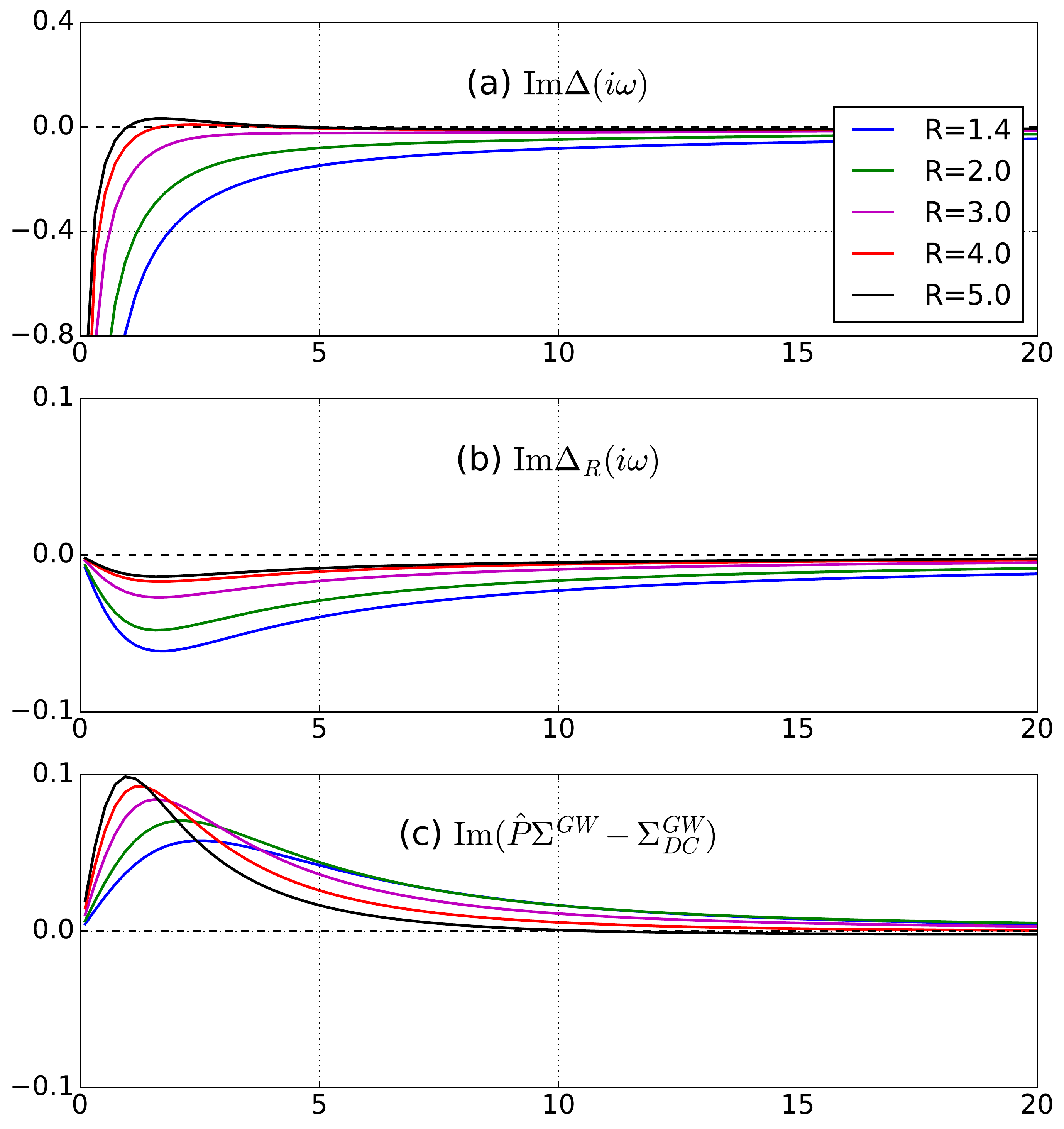}
\caption{(Color online) The imaginary part of (a) the hybridization function $\Delta(\iw)$, and its component (b) $\Delta_{R}(\iw)$ and  (c) $\Delta\Sigma^{GW}_{DC}(\iw)$of Eq.~\eqref{hyb:res}. For the two cases (R=4.0 and 5.0) in which the causality break down, we took the result of the first iteration.}
\label{fig4}
\end{figure}
In Fig.~\ref{fig4}(a) we show the imaginary part of $\Delta$ on the
imaginary frequency axis. It is clear that for $R\gtrsim 3.5$ the
imaginary part of $\Delta$ becomes positive in some frequency regime,
violating the causality. In Figs.~\ref{fig4}(b) and (c), we also present
the terms appearing in Eq.~\ref{hyb:res}, i.e.,
$\Delta_{R}$ and $(\hat P \tilde \Sigma^{GW}-\tilde \Sigma^{GW}_{DC})$. 
Clearly $\Delta_R$ is always causal, while $(\hat P
\tilde \Sigma^{GW}-\tilde \Sigma^{GW}_{DC})$ has the wrong sign. 
In the weakly to moderately correlated regime, the last term in the denominator of \eqref{Eq:47} is large since the hopping between the two DMFT orbital, $S_{12}$, is substantial as the two site are close to each other. Therefore, this term outweighs the non-causal term $(\hat P \tilde \Sigma^{GW}-\tilde \Sigma^{GW}_{DC})$ and the causality is not yet violated in the weakly correlated regime.

One might ask then how is such causality violation avoided in the
exact solution, .i.e., when we replace $\Sigma^{GW}$ with sum of all
non-local Feynman diagrams. We know that in this case we should
recover the exact solution. In this particular case, $\Sigma_{DC}$
would vanish, as all terms are non-local and thus nothing is
double-counted. We would then need to see that the projection of the
non-local diagrams to the correlated site is positive. But there is a
second possibility, which is more likely in low dimensional systems
and molecules, namely that the non-local part of the self-energy
diverges simultaneously with the local part, and therefore the
separation into diverging local and well-behaved non-local part is not
possible. In another words, we would not be able to neglect the last
term in the denominator Eq.~\ref{Eq:47}, because $S_{12}$ is as large
as $\Sigma_{imp}$. In $GW$+DMFT, the $GW$ self-energy is always
Fermi-liquid like and it never diverges.

\begin{figure}[h!]
\includegraphics[width=1.0\linewidth,clip=]{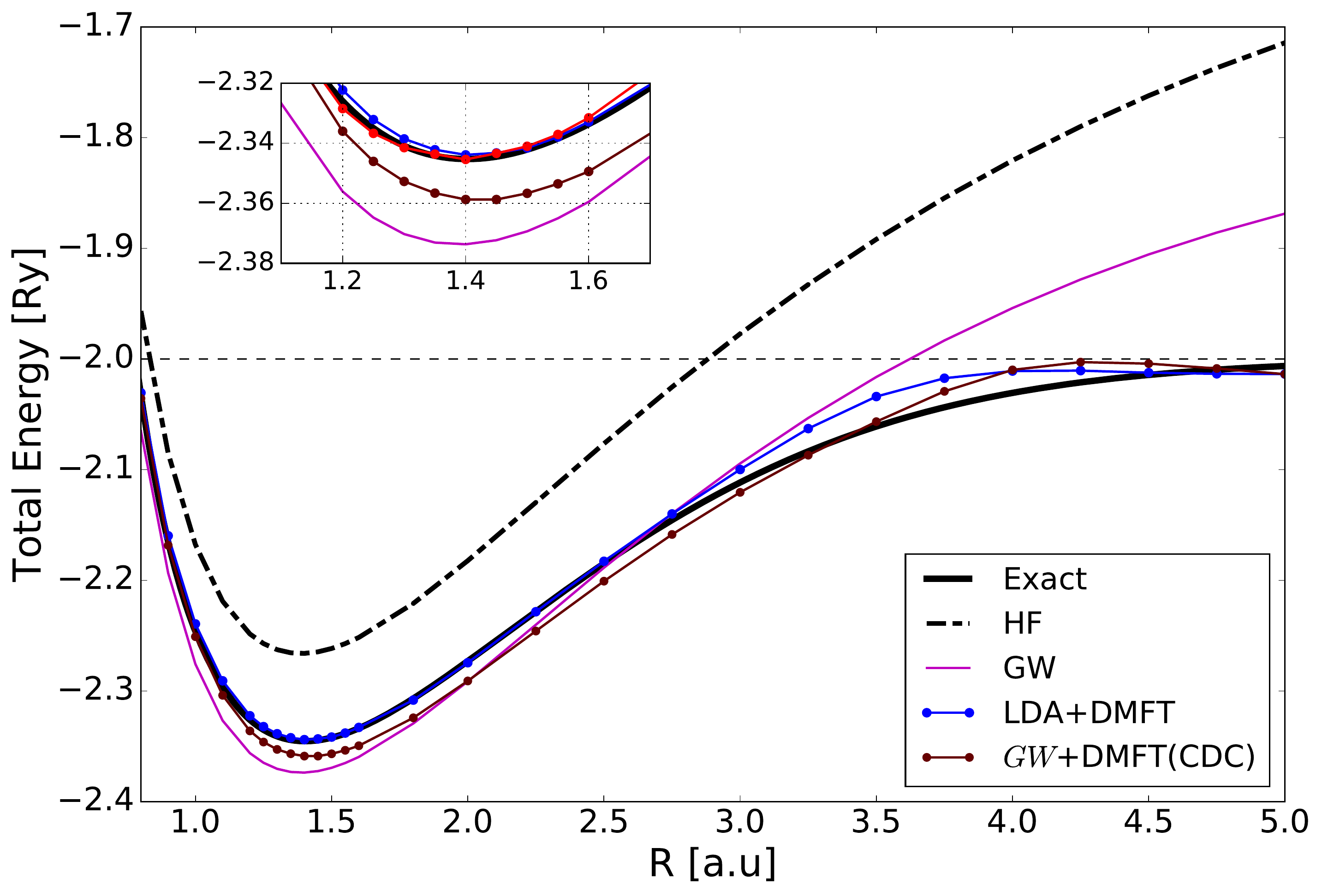}
\caption{The total energy result of GW+DMFT with causal double-counting (CDC) scheme}
\label{fig_ER_causal}
\end{figure}

%@@@@
As clear from the above argument, the causality is not violated 
if we take the CDC (Eq.~\eqref{CDC}) because $P\hat \Sigma^{GW}-\Sigma^{DC}=0$.
We present the total energy of this scheme in the Fig.~\ref{fig_ER_causal}. 

In the moderate correlated regime (around $R=1.4$), the CDC scheme is worse than exact double-counting scheme both in total energy and spectral function. The error of the total energy of GW+DMFT with CDC, 0.8\%, is worse than that of LDA+DMFT (0.2\%) or GW+DMFT with exact DC (0.3\%). However, it is very important that the CDC double-counting works correctly at the large distance regime, where all the exact double-counting schemes fail due to the causality violation. As clearly seen, the CDC total energy converges to -2.0.

Although CDC is an ad-hoc scheme, it allows $GW$+DMFT to work in the strongly correlated regime without violating the causality. Since GW+DMFT is meant for strongly correlated solids where $Z$ is typically small, we argue that the $GW$+DMFT for solid-state calculation is better when  the CDC scheme is used and the effect of using the CDC instead of the exact double-counting should be investigated systematically.

\subsubsection{Spectral function}
In this section, we present the spectral function, the imaginary part
of the Green's function summed over all diagonal component. We choose
the zero of energy corresponding to dissociation of the molecule, so
that the negative peak in spectra corresponds to the ionization
energy.
The Pad\'{e} method is used for analytic continuation from imaginary
frequency to real frequency. We mention in passing that Pad\'{e}
approximation is very accurate here, and we found only very minor
sensitivity of the pole position (around 0.1\%) depending on the
choice of the input parameter for Pad\'{e} method.

% 
% The chemical potential is
% determined in such a way that the total number of electrons is 2.0.
% The occupied spectrum (the lowest energy peak) then represents the
% negative ionization energy ($-\mathrm{IE} = E_{H_2}-E_{H_2^{+}}$).

%As expected, for $R=1.4$, the lowest unoccupied peak (electronic affinity) is ends up above zero so we will only consider the occupied peak (the negative ionization energy). At $R=5.0$ where the two atoms are fairly separated, the orbital hybridization is weak enough for the lowest two orbitals to be nearly degenerate. As a result, we can clearly observe in this case two $-$IE lines that are quite close to each other. \\

\begin{figure}[h!]
\centering{
\includegraphics[width=1.0\linewidth,clip=]{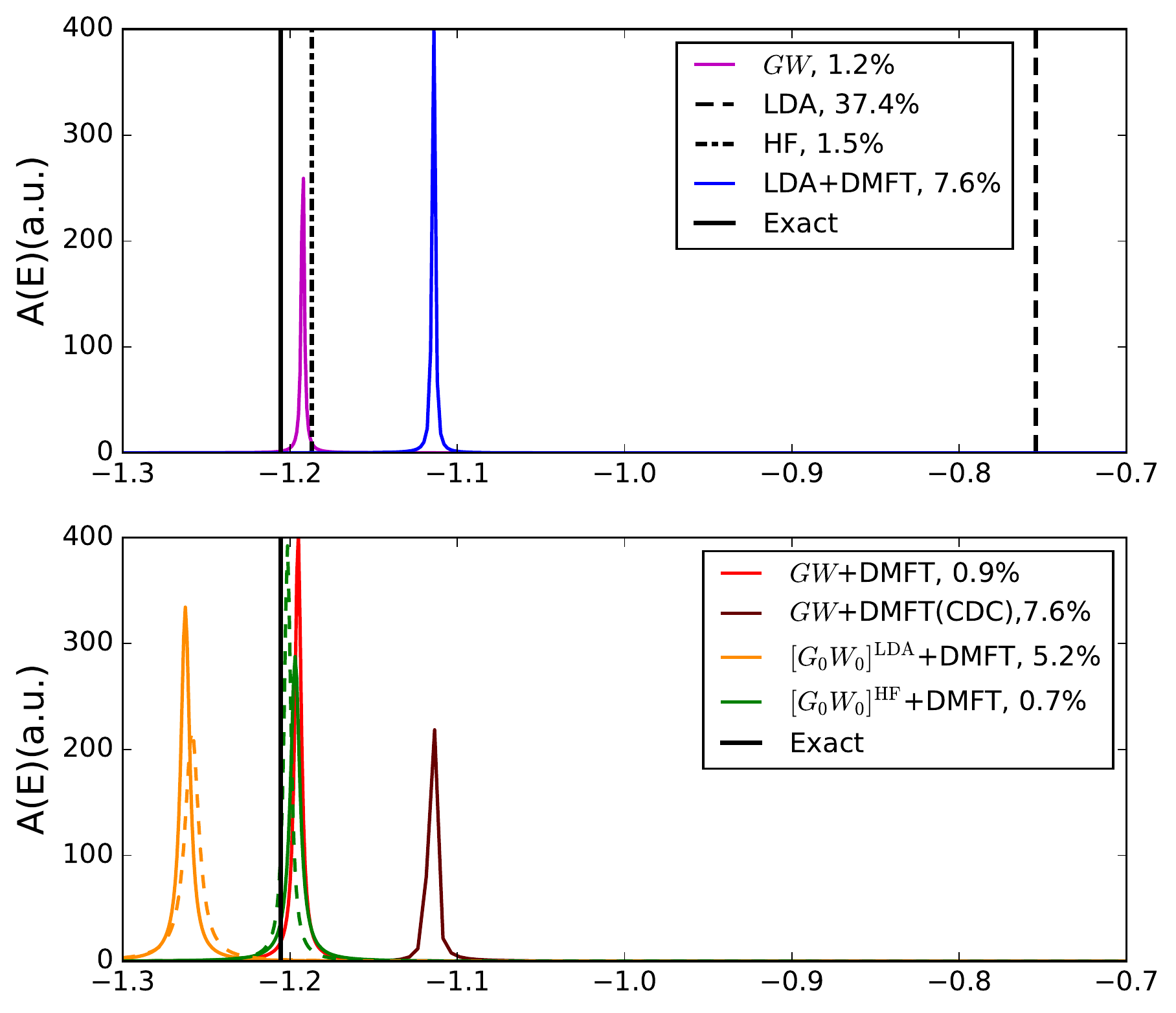}
}
\caption{(Color online) Spectral function for $R=1.4$ where the exact IE is $-1.206~a.u$. Each errorbar is presented in the legend. (\emph{upper panel}) $GW$, LDA, HF and LDA+DMFT for comparison. (\emph{lower panel}) $GW$+DMFT schemes with different self-consistent conditions. The dashed lines indicates the result without charge self-consistency}
\label{spectral_gwd}
\end{figure}
In Fig.~\ref{spectral_gwd} we display the spectral function for $R=1.4$ (the equilibrium
distance), which corresponds to the moderately correlated regime. The
peak position, measured from the vacuum (not from the chemical potential)
corresponds to the IE.
The LDA prediction for IE is 40\% off the exact value. This failure
is related to the band gap underestimation in solid-state calculation.
On the contrary, the HF prediction is very good in this regime, and is
only 1.5\% off the exact value. The $GW$ approximation slightly
improves on HF, and its IE in the moderately correlated regime is only
1.2\% of the exact value.

In Fig.~\ref{spectral_gwd}a we also show the LDA+DMFT prediction,
which substantially improves the LDA value from 40\% error down to
7.6\% error. Nevertheless, the LDA+DMFT spectra is not very accurate,
as it builds on too inaccurate starting spectra. 

We expect that GW in combination with DMFT improves the GW result.
Indeed when combining GW and DMFT in a fully self-consistent way, the
error is only 0.9\% and when using $G_0W_0$ from HF (which itself is
quite precise), the error is only 0.7\%. Somewhat worse is the result
of $G_0W_0$+DMFT when $G_0$ is taken from LDA. The error in this case
is quite comparable to LDA+DMFT error, but it seems the combination of
DMFT and $G_0W_0$ overcorrects the LDA. 

Notice also that the charge self-consistency has almost no effect on
the spectra, while we showed before that charge self-consistency is crucial
for the accuracy of the total energy. On the other hand, the choice of
$G_0$ is crucial for spectra, but not for the total energy.

%@@@@@ as kristjan edited.
We note that the GW+DMFT with CDC is worse than . We attribute
this inaccuracy to the fact that the CDC scheme includes the interactions between
the DMFT space and the rest of Hilbert space, 
which is not supposed to be involved in the local impurity self-energy.
Therefore, CDC scheme in this scheme is less precise than the $GW$+DMFT
with the exact DC.
 
\begin{figure}[h!]
\includegraphics[width=1.0\linewidth,clip=]{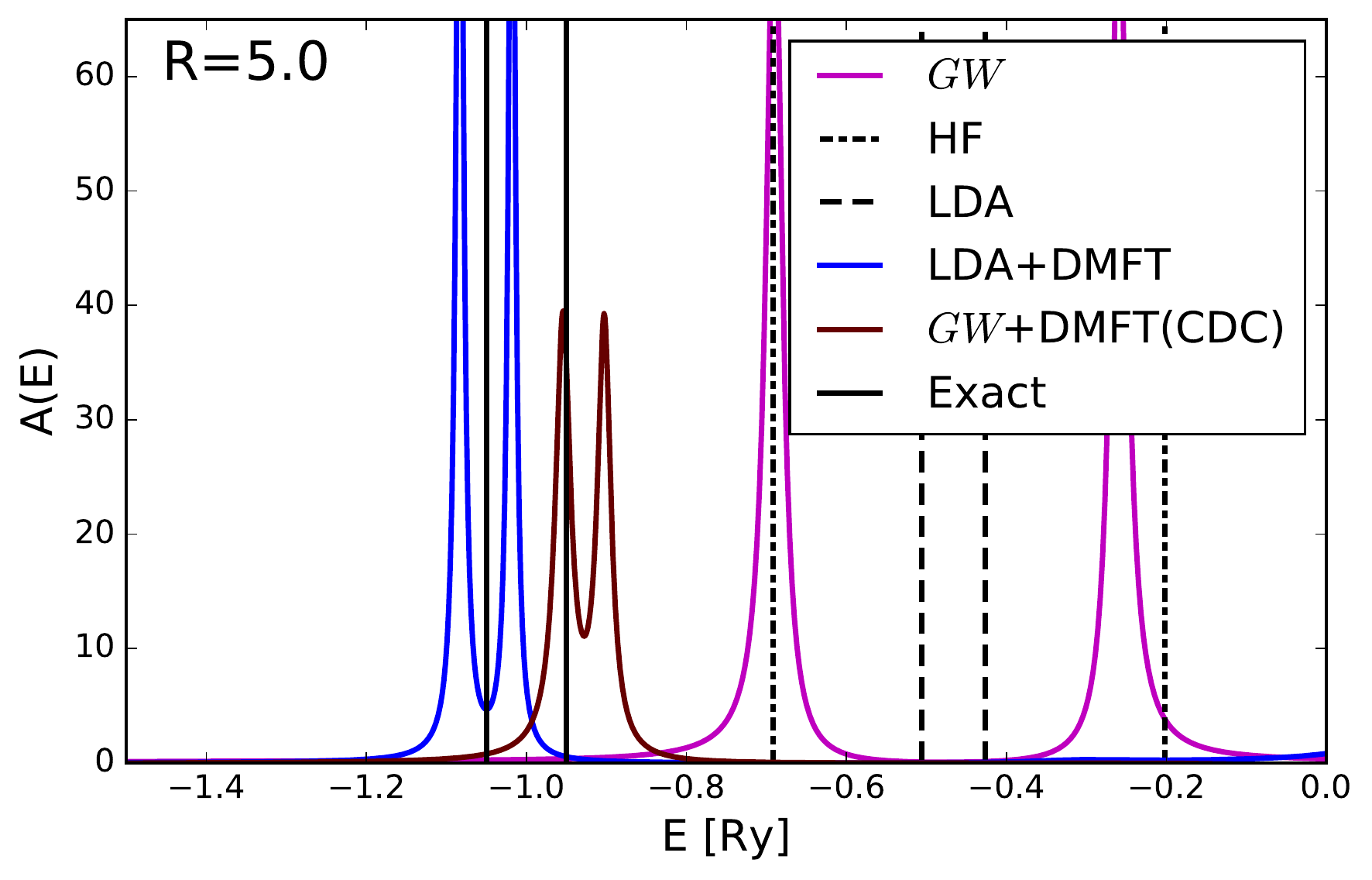}
\caption{The spectral function (\emph{upper panel} R=1.4, \emph{upper panel} R=5.0) calculated by GW+DMFT with causal double-counting (CDC) scheme}
\label{fig_spec_causal}
\end{figure}

On the other hand, at the large separation limit (Fig.~\ref{fig_spec_causal})
where the correlation effects are strong,
the CDC scheme is the only $GW$+DMFT method that works without causality violation.
It successfully reproduces the spectral function close to the exact value and its
quality is comparable to that of LDA+DMFT.
All other methods without the DMFT treatment, LDA, HF and $GW$, are far from the exact ones due to 
the failure of perturbation theory.

We notice that DMFT considerably improves the total energy of $GW$ in
this regime, while the spectra seems barely affected. This is because
the renormalization amplitude from local GW self-energy and the DMFT
self-energy are almost the same in this weakly correlated regime,
and their values are 0.935 and 0.928, respectively.

\subsection{QS$GW$+DMFT}

As shown above, the self-consistent $GW$+DMFT fails in the correlated
regime due to causality violation, which comes from the fact that
double-counted self-energy is dynamic and too large. In QS$GW$ the GW
spectra is represented by an approximate static Hermitian Hamiltonian,
and in this case we expect that approximating the double-counting by a
static value might be a reasonable choice. We denoted this choice by
SDC. As discussed above, the static DC term tends to over-count the
renormalization effects, and this can be somewhat remediet by choosing
dynamic double counting, which we denote by DDC.

%\subsubsection{Total energy}

%%%%%%%%%%%%%%%%%%%%%%%%%
%We present the QS$GW$+DMFT total energy results with two different
%double counting schemes in 

\begin{figure}[h!]
\centering{
\includegraphics[width=1.0\linewidth,clip=]{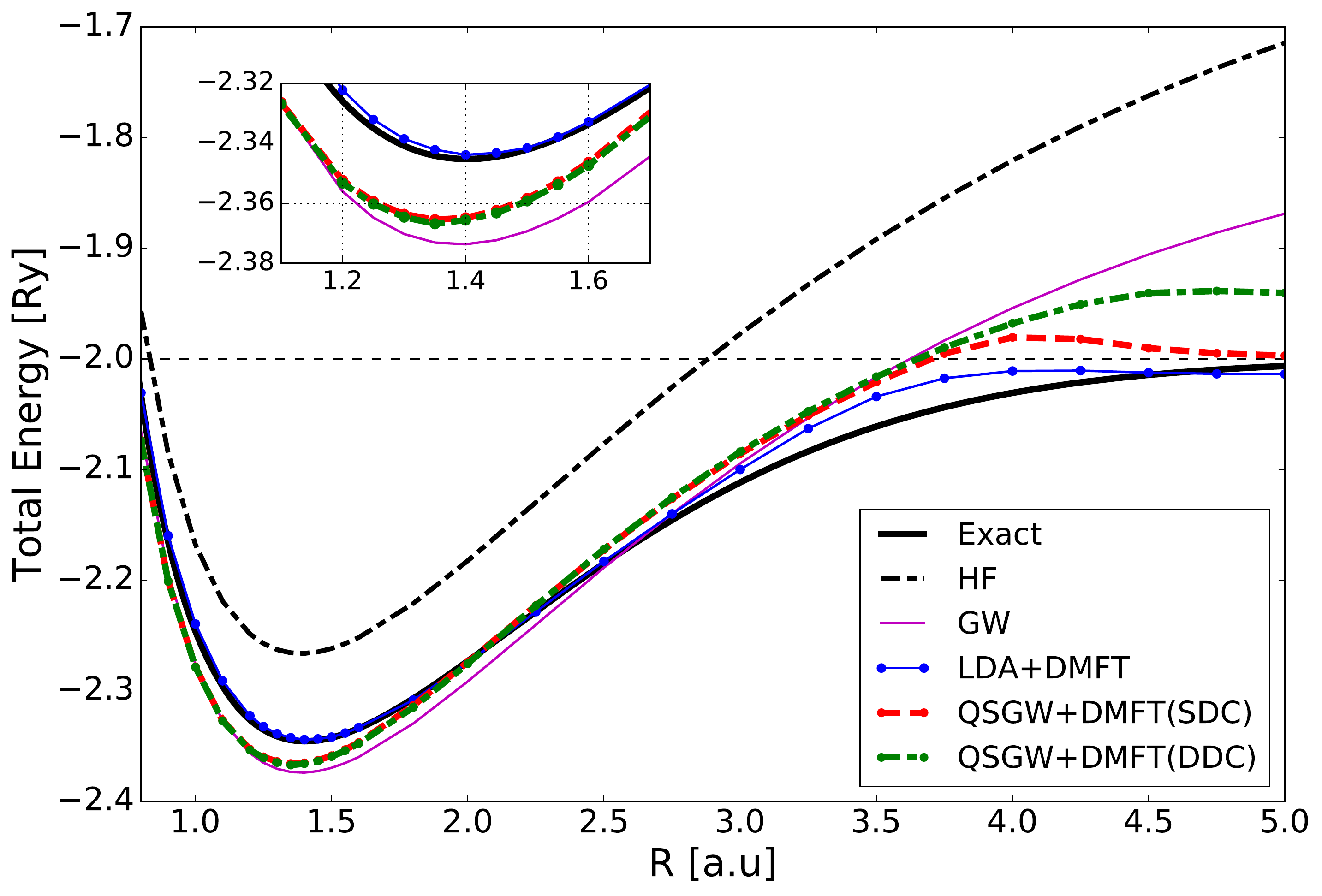}
}
 \caption{(Color online) The ground state energy of the QS$GW$+DMFT results with two different schemes of double-counting. The inset shows the magnification around the equilibrium distance.} \label{energy_qsd}
\end{figure}
In Figs.~\ref{energy_qsd} we display the total energy of QS$GW$+DMFT
together with $GW$,  LDA+DMFT and exact total energy. Because the
quasiparticle approximation is not derivable from a functional, the
energies are unfortunately not very good. At equilibrium, the
QS$GW$+DMFT energy is very similar to $GW$ energy, while in the
correlated regime the addition of DMFT slightly improves the $GW$ energy.
Nevertheless, the dynamic double-counting does not recover correct
atomic limit, even though DMFT is expected to be exact in the atomic
limit. This failure is again due to the double-counting issue, namely,
when impurity self-energy is diverging, the hybridization should
vanish, but when a dynamic double-counting is used, hybridization does
not vanish, and the atomic limit is not reached. In solid state
applications, this would correspond to a missed Mott transition in
the strongly correlated limit. We see in Fig.~\ref{energy_qsd} that only a
static DC correctly reproduces atomic limit. But unfortunately the
total energy is substantially worse than corresponding LDA+DMFT
result. This is not unexpected, as only methods derivable from a
functional are expected to give precise total
energies.~\cite{leeuwan2006epl}

\begin{figure}[h!]
\centering{
\includegraphics[width=1.0\linewidth,clip=]{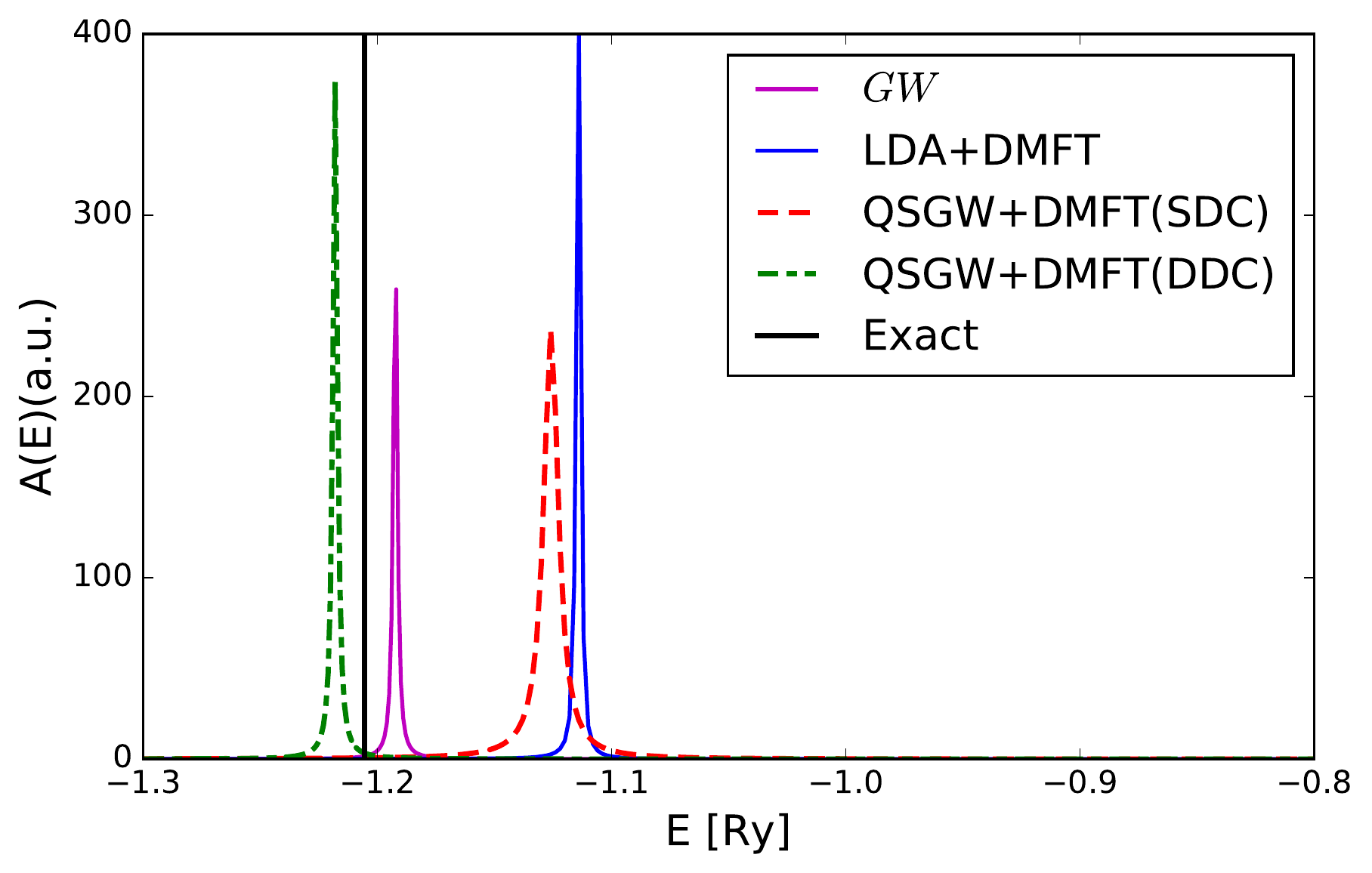}}
\caption{(Color online) The QS$GW$+DMFT spectral results with two different double-counting schemes at $R=1.4$, compared with GW, LDA, HF and LDA+DMFT}
\label{spectral_qsd1.4}
\end{figure}

Next we show the spectral functions of QS$GW$+DMFT at equilibrium
position. Fig.~\ref{spectral_qsd1.4} compares the GW, LDA+DMFT and two
version of QS$GW$+DMFT schemes with the exact solution. For the static
double-counting (SDC) scheme, the spectra is not good, and very
comparable to LDA+DMFT result. The origin of the error is however
quite different, in QS$GW$ it is due to the double renormalization by
both the $GW$ and DMFT, while in LDA+DMFT it is due to missing non-local
correlations.

 The dynamic double-counting scheme (DDC) substantially improves the
spectra in this weakly correlated regime, and the error of IE is only
1\%, comparable to the fully self consistent GW+DMFT.
However, the DDC scheme is much worse in the strongly correlated
regime, both for energy and for spectra. 

\begin{figure}[h!] \centering{
    \includegraphics[width=1.0\linewidth,clip=]{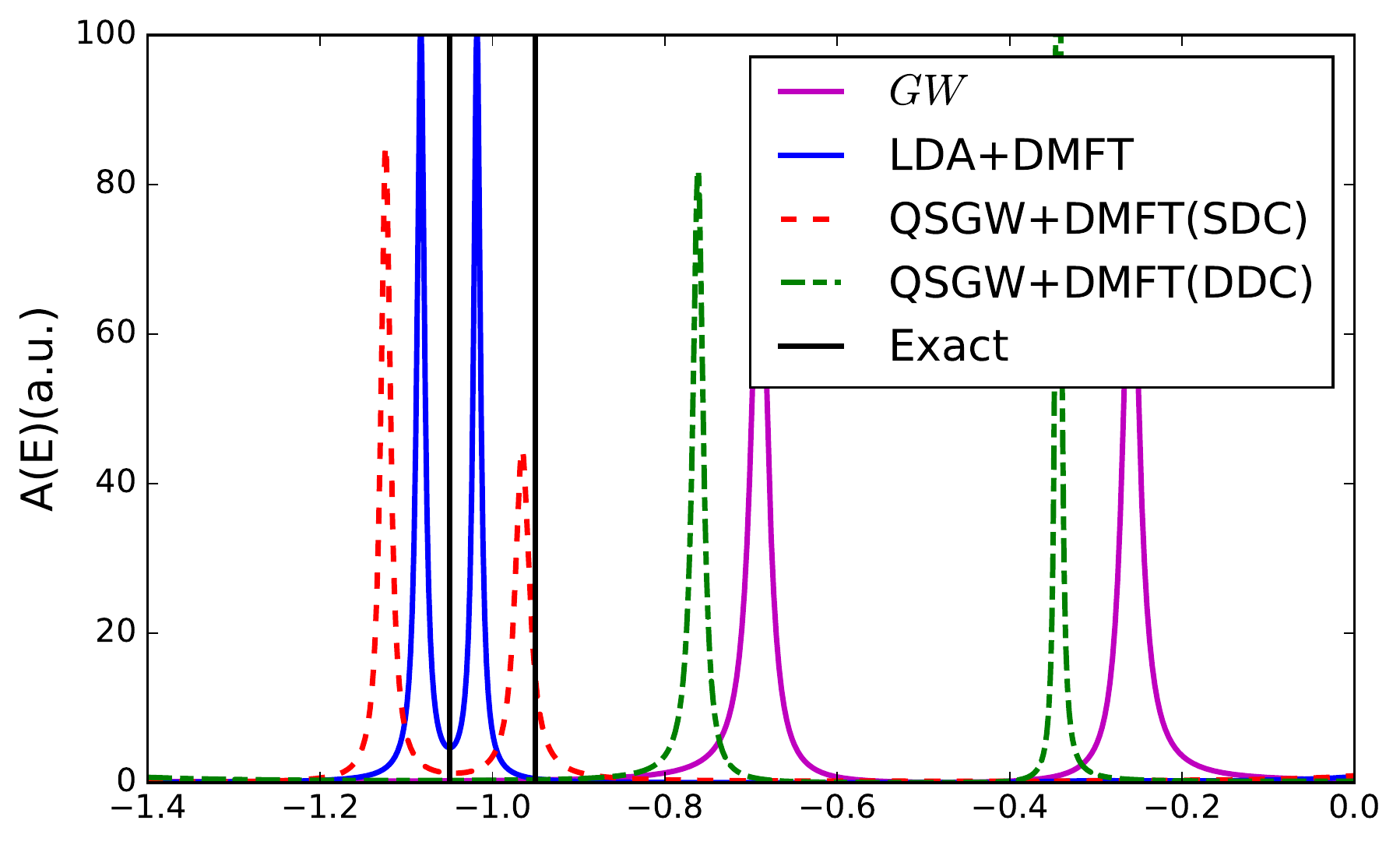} }
  \caption{(Color online) Spectral function for $R=5.0$. (\emph{upper
      panel}) Perturbative schemes: HF, LDA and $GW$. (\emph{lower
      panel}) LDA+DMFT and QS$GW$+DMFT with two different
    double-counting functional.} \label{spectral_qsd5.0} 
\end{figure}
 In Figs.~\ref{spectral_qsd5.0} we present results at $R=5.0$, deep in
the correlated regime, where excitations of two almost independent H atoms
are expected, with the value close to $-1.0\,$Ry.
In this regime QS$GW$+DMFT with static double-counting (SDC) and
LDA+DMFT perform reasonably well, while dynamic double-counting (DDC)
fails very similarly to $GW$ approximation.
This failure of DDC was also reported in Ref.~\cite{sakuma2013prb},
where a similar scheme to our DDC was tested on SrVO$_3$.

\section{Summary}

We have implemented $GW$+DMFT and QS$GW$+DMFT scheme for H$_2$
molecule, and we compared the total energy and spectral function with
the exact result, and LDA+DMFT. For $GW$+DMFT, five different
calculations have been performed: (i) fully self-consistent $GW$+DMFT,
(ii) $[G_0W_0]^{\mathrm{HF}}$+DMFT where $G_0$ is taken from Hartree-Fock,
(iii)$[G_0W_0]^{\mathrm{LDA}}$+DMFT with $G_0$ from LDA, (iv)
$[G_0W_0]^{\mathrm{HF}}$+DMFT but without charge self-consistency, 
(v) $[G_0W_0]^{\mathrm{LDA}}$+DMFT without charge self-consistency.

1) In the strongly correlated regime only LDA+DMFT and QSGW+DMFT with
static double-counting give good spectra, and only LDA+DMFT gives good
total energy.

2) Most of $GW$+DMFT schemes fail in the correlated regime due to
causality violation. While QS$GW$+DMFT does not suffer causality
violation, it performs reasonably well in the correlated regime only
when using the static double-counting.

3) In the Fermi liquid regime of weak to moderate correlations, fully
self-consistent $GW$+DMFT is excellent, both for total energy and
spectra.

4) The spectra in the weakly correlated regime is also very accurately
obtained by $[G_0W_0]^{\mathrm{HF}}$+DMFT, but less precise
with $[G_0W_0]^{\mathrm{LDA}}$+DMFT. The QS$GW$+DMFT with static
double-counting, which performs well in correlated regime, is less
precise here, as it renormalizes spectra twice. The dynamic
double-counting remedies this shortcoming in the weakly correlated regime,
but fails in the strongly correlated regime.

5) Total energy in the weakly correlated regime is good in all
$GW$+DMFT schemes (but not in QS$GW$+DMFT), provided the charge is
computed self-consistently.

In summary, the strongly correlated regime is more challenging to
describe by $GW$+DMFT as previously thought, and the causality
violation seriously impacts the prospects of using $GW$+DMFT in this
regime. On the other hand, using QS$GW$+DMFT in this regime leads to
somewhat better spectra than employing less demanding LDA+DMFT, but it
does not lead to better total energies.

\section{Acknowledgments}
J.L. thanks  Lorenzo Sponza for discussion of the causality issue. 
J.L. was supported by the Rutgers Physics Departmental Fellowship and NSF-DMR 1405303.
K.H. was supported by the NSF-DMR 1405303 and the Simons foundation
under project ``Many Electron Problem".

\bibliography{../../mybib}

\end{document}